\documentclass[acmsmall]{acmart}
\AtBeginDocument{%
  }

\setcopyright{acmlicensed}
\copyrightyear{2024}
\acmYear{2024}
\acmDOI{XXXXXXX.XXXXXXX}

\acmJournal{TORS}
\acmVolume{1}
\acmNumber{1}
\acmArticle{1}
\acmMonth{1}

\usepackage{booktabs}
\usepackage{tabularx}

\usepackage{todonotes}
\presetkeys%
    {todonotes}%
    {inline,backgroundcolor=yellow}{}

\begin{document}

\title[Rethinking Group Recommender Systems in the Era of GenAI]{Rethinking Group Recommender Systems in the Era of Generative AI: From One-Shot Recommendations to 
Conversational Group Decision Support}

\author{Dietmar Jannach}
\email{dietmar.jannach@aau.at}
\orcid{0000-0002-4698-8507}
\affiliation{%
  \institution{University of Klagenfurt}
  \city{Klagenfurt}
  \country{Austria}
}

\author{Amra Delić}
\email{adelic@etf.unsa.ba}
\orcid{0000-0002-5584-6893}
\affiliation{%
  \institution{University of Sarajevo}
  \country{Bosnia }
}

\author{Francesco Ricci}
\email{fricci@unibz.it}
\orcid{0000-0001-5931-5566}
\affiliation{%
  \institution{Free University of Bozen-Bolzano}
  \country{Italy}
}

\author{Markus Zanker}
\email{markus.zanker@acm.org}
\orcid{0000-0002-4805-5516}
\affiliation{%
  \institution{Free University of Bozen-Bolzano and University of Klagenfurt}
  \country{Italy and Austria }}

\begin{abstract}
More than twenty-five years ago, early ideas emerged on how to design systems that provide recommendations to groups of users rather than individuals. Since then, a wide range of algorithmic approaches has been proposed to elicit individual preferences, aggregate them, and generate group recommendations. However, despite this substantial body of work, relatively few examples of real-world group recommender systems are in use. This observation invites reflection on the assumptions of the academic research, particularly regarding how inter-group communication unfolds and how recommendation-supported decisions should be made. In this essay, we suggest that some of these assumptions and corresponding system designs may not fully align with users' needs or expectations. We therefore outline a forward-looking perspective that explores how insights from group-decision making research and recent advances in Generative AI, including systems such as ChatGPT, could better inform the design of group recommender systems. As one possible direction, we consider settings in which human group members interact and exchange information through a chat based interface, while an AI-based agent supports the decision-making process. Such an approach may contribute to design more natural group and system interactions, and could help facilitate a broader adoption of group recommender systems in practice.

\end{abstract}



\maketitle

\section{Introduction}
\label{sec:introduction}
The idea of making automated recommendations to \emph{groups of users} instead of targeting the preferences of \emph{individuals} has been around for at least twenty-five years. In an early work from 1998, McCarthy and Anagnost~\cite{McCarthy1998MusicFX} proposed MusicFX, a ``group preference arbitration'' system, which allows members of a fitness center to influence the music that is played in this shared environment. In that system, fitness center members could enter their genre preferences, and MusicFX then selects appropriate radio stations to play, i.e., depending on the preferences of the currently present crowd. PolyLens is another early example of a \emph{group recommender system} (GRS)~\cite{OConnor2001PolyLens}, which was deployed as an extension to the MovieLens recommender system in the year 2000. With PolyLens, MovieLens users can create new groups by inviting other members. In a second step, individually expressed preferences of group members are aggregated by PolyLens to make recommendations to the group.

A shared aspect of these two early systems is that they have been deployed in real-world environments, one in a physical space, the other in the online sphere. In both cases, survey studies after several weeks of deployment indicated that the respective systems were well appreciated by the users. In addition, in both works, various relevant questions beyond the quality perception of the recommendations are discussed, e.g., regarding how users interact with the system, how they express preferences, how groups are formed, how users think about privacy aspects, or how the user experience should be designed. Furthermore, in the case of MusicFX, questions of fairness and diversity were addressed as well, for instance, to avoid monotony that arises from focusing on the least common denominator regarding musical tastes or to avoid ``starvation'' of minority tastes.

\paragraph{Today's Research Focus.}
Around twenty-five years later, the topic of group recommender systems is well-established within the broader recommender systems literature. The state-of-the-art in this area is well-documented in various overview works such as books, handbook chapters, and survey papers~\cite{Jameson2007,Felfernig2018GRSIntro,Masthoff2022GRSBeyond,Masthoff2015GRS,Singhal2024SOTS,Dara2019Survey}.
At the same time, these overviews tend to place substantial emphasis on algorithmic aspects, particularly methods for aggregating individual preferences and to generate group recommendations. Other aspects -- such as preference elicitation, explanation, and support for group decision-making -- receive comparatively less attention or are treated as outside the primary scope~\cite{Masthoff2022GRSBeyond}. Similarly, although the importance of user-centered evaluation is frequently acknowledged, discussions of evaluation methodologies focus predominantly on offline experimental settings without direct user involvement~\cite{Felfernig2018GRSIntro}.

Another observation is that relatively few recent examples describe group recommender systems deployed in real-world environments, even as prototypes, in contrast to some of the earlier work. While survey papers do highlight a range of systems and application scenarios (e.g.,~\cite{Masthoff2022GRSBeyond,Singhal2024SOTS}), only a limited number appear to have been evaluated in naturalistic settings. Moreover, group recommendation solutions are not yet widely visible on major platforms such as Netflix, YouTube, or Spotify. There are, of course, several studies that involve user evaluations~\cite{McCarthy2006Needs,Yu2006TV,Felfernig2012Requirements,Guzzi2011Interactive,alvarez2018negotiation,Najafian2021Factors,Delic2018Observational}, however, these are often conducted in controlled environments using research prototypes.

These observations invite reflection on whether current academic research in group recommender systems is addressing the most relevant questions and usage scenarios. For instance, it remains unclear how often and when people would engage with a dedicated group recommendation tool -- such as for movies -- by explicitly specifying their preferences and ratings, and then relying primarily on a system generated ranked list, algorithmically determined by the system on the base of the aggregated inputs. In practice, it may be more convenient for groups of friends to use existing communication tools, such as messenger apps, to discuss or negotiate which movie to watch together on a weekend, see~\cite{Church2013,Kim2020Bot,DelicE0M24}.

\paragraph{On Conversational Group Decision Making.}
It is certainly not a new observation that group decision-making in reality is often a highly interactive process\footnote{Already in 2007, Jameson and Smyth critically~\cite{Jameson2007} mention the problem of limited support in existing GRS for achievement of consensus.}. An earlier work in that direction, Guzzi et al.~\cite{Guzzi2011Interactive}, presented the \emph{where2eat} GRS prototype that leverages an interactive critiquing-based method where users exchange recommendations among each other, supported by a recommender system that helps retrieve suitable alternatives. More recently, Álvarez-Márquez and Ziegler~\cite{alvarez2018negotiation} proposed an approach that focuses on the group's social interactions during the discussion and negotiation in the decision-making process. A similar approach was taken later by Contreras et al.~\cite{Contreras2021Integrating}, who proposed a \emph{conversational} group recommendation approach which also takes into account the different \emph{roles} of the group members, i.e., leaders or collaborative users, in such a conversation.

A main assumption of such conversational approaches is that a group recommender system should rather act as a \emph{facilitator} of the group's decision-making process than as a normative judge that asserts which options merit the group attention. In other words, the main bulk of the communication flow in the decision-making process should happen \emph{among} the group members, and not primarily between the users and the GRS~\cite{Jameson2022}. Therefore, when group members, for example, express their preferences, the main goal of communication should \emph{not} be to inform the GRS, but to share information with the group. Thus, the user interface of a GRS must provide appropriate means to support user-to-user communication. One central way to enable such a communication is the provision of a \emph{chat} functionality, where users can interact in natural language in real-time. In some approaches like~\cite{alvarez2018negotiation}, the system's user interface therefore not only has GRS-specific UI elements, e.g., to explore item features, group preferences, or recommendations, but also an integrated chat interface. 

Instead of relying solely on a chat feature as \emph{one of several elements} of the GRS user interface as in~\cite{alvarez2018negotiation}, Nguyen and Ricci~\cite{Nguyen2017AChatBased} propose a tourism recommender system where the group chat functionality takes a central place in the app. Later on, Delic et al.~\cite{Delic2023Charm} developed this idea further and proposed a framework for group recommendations (CHARM), which implements a chatbot that can be integrated into existing chat platforms like WhatsApp or Telegram. The goal of CHARM is to mediate the decision-making process by providing certain types of information \emph{upon request}, e.g., summarize the items discussed, make a recommendation, or create a choice set from items that were positively assessed in the current session. Overall, by proposing the CHARM framework, Delic et al.~argue for a new approach for group recommendation, which, as noted above, goes beyond the questions of preference aggregation of individual group members, and which instead supports group-decision making in a more \emph{holistic} way. In their concluding remarks, Delic et al.~envision a future version of the framework that automatically detects user intents, rather than users being forced to state their intents, through Natural Language Processing (NLP) techniques. First steps towards the implementation of such a framework are laid out in~\cite{Karahodza2025GroupDynamics}. Furthermore, they mention the potential use of Generative AI\footnote{In the context of chat-based approaches, the term Generative AI is often used interchangeably with the term LLM.}, but remain skeptical that central functionalities of a GRS, e.g., identifying the status of the decision-making process, can be implemented with such technologies.

\paragraph{Towards AI-Supported Conversational Group Recommendation.}
In this essay, we embrace the ideas put forward by Delic et al.~\cite{Delic2023Charm} towards a new generation of conversational group recommender systems as facilitators of decision-making processes. We, however, take a more optimistic perspective on the use of Generative AI in general, and Large Language Models (LLMs) in particular, as facilitators of next-generation chat-based GRS. Specifically, we believe that Generative AI techniques offer a full range of new opportunities for future GRS. Existing research, for example, shows that LLMs can be directly used as zero-shot or few-shot recommender systems in certain use cases, without the need of involving additional item ranking algorithms~\cite{hou2024large}. Furthermore, tools like ChatGPT have been successfully explored in conversational recommendation settings, where they are able to understand user intents with impressive accuracy or provide explanations to users~\cite{he2023large,friedman2023leveraginglargelanguagemodels,feng2023largelanguagemodelenhanced}. LLMs can also be excellent for NLP tasks such as text summarization, which can be a desirable and highly valuable feature to support group discussions.

In fact, recent literature provides ample evidence for the effectiveness of LLMs in conversational recommendation scenarios with a single user~\cite{Wagne2025CRS, Wu2024ASurvey, Manzoor2024Chatgpt}, and our envisioned system shares many commonalities with conversational recommender systems (CRS) in terms of text understanding and text generation.\footnote{What sets our vision apart from single-user CRS is the support for group coordination and group decision-making. }
Furthermore, we believe that modern LLMs are able to facilitate decision processes at higher levels in a multi-faceted way, e.g., by monitoring the behavior of groups, identify members who have not been heard, proactively stimulate individual contributions or even by deescalating recognized conflicts. This way, LLM-based recommendation agents will be able to fulfill various roles in group discussion as outlined recently by Ricci et al.~in~\cite{ricci2025wideningrolegrouprecommender}.

\paragraph{Building Carefully on Strong Foundations.}
Our vision builds on two complementary foundations. First, research on human decision-making and group decision-making has produced a rich body of insights into how people exchange information, negotiate preferences, reach consensus, and coordinate their decisions. Second, recent advances in Generative AI provide new technical capabilities for supporting such processes in more flexible, interactive, and context-aware ways. We therefore adopt a dual perspective: rather than deriving our vision primarily from what new AI technologies make possible, we aim to connect these emerging capabilities with established insights into how human groups actually make decisions.

At this point, we emphasize that the goal of this essay is not to diminish the importance of prior work and criticize their contribution, but to motivate a forward-looking vision that builds upon it. Specifically, we reflect on how group recommender systems can be seamless integrated into everyday usage contexts and decision-making processes, rather than suggesting deficiencies in existing approaches. For example, established methods for preference aggregation are still worth to be incorporated into the envisioned framework, and we strongly advocate that the design of any AI-based decision-support agent should be rooted on the many compelling insights that the literature has produced. We also acknowledge that our vision -- conceived as a \emph{class of systems} rather than a single concrete solution -- is targeted to certain decision-making scenarios where group members are usually investing time in setting the decision goal, proposing options and discussing how their preferences are matched by the alternative options. More mechanical and less interactive use cases, such as music selection in a gym setting, may be better addressed by alternative and simpler approaches. Furthermore, we are aware of various potential limitations and risks of Generative AI, as will be outlined later in this essay.

In the remainder of this \emph{perspectives} paper, we first review existing insights on human-decision processes in recommender systems, chat-based decision support and conversational group decision-making in Section~\ref{sec:background}. Then, we outline our vision of future group recommender systems based on Generative AI in the form of LLMs in more depth and sketch a possible research agenda in Section~\ref{sec:research-agenda}.

\section{Understanding Decision-Making Processes}
\label{sec:background}
In this section, we review three complementary strands of research that provide foundations for our vision of next-generation group recommender systems. We first discuss insights from research on human decision-making in recommender systems, with a particular focus on phenomena that arise in group settings. We then review existing chat-based group recommender systems and, finally, broaden the perspective to conversational agents for group decision support beyond the recommender systems literature. Together, these strands provide insights into both, how groups make decisions and how computational agents can support these processes, thereby informing the vision of LLM-supported group recommendation presented in Section~\ref{sec:research-agenda}.
\subsection{Human Decision-Making in Group Recommender Systems}
\label{sec:human-decision-making}
A considerable part of the literature on recommender systems is dedicated to the design of algorithms for personalized item rankings. How users actually make choices when provided with a list of item suggestions and how a recommender system can support decision-making processes is unfortunately much less explored~\cite{Chen2013Human}. In many application use cases, decision-making may often follow a mostly rational process, for instance, when users compare item features in the light of their own needs and preferences. However, there are also many other factors that may influence the decision-making processes of individual users, among them various psychological phenomena like cognitive biases, see~\cite{schedl2024importance,Lex2021Psy} for recent surveys. In group recommendation scenarios, even more factors can affect the interaction and choice behavior of group members, for example when they strive for conformity to the group opinion~\cite{Forsyth2019,MasthoffG06}. Thus, it is highly important to understand such phenomena when aiming to build an effective group recommender system.

\paragraph{From Reducing the Choice Set to High-Quality Decisions}
\citet{Jameson2022}~review questions of human decision-making in recommender systems, both for individual user decision scenarios and for decisions in group settings. They begin their analysis with a discussion of the possible purposes of recommender systems. They find that in some cases the purpose of the system is mainly to \emph{reduce the choice set} through information filtering, whereas in other cases the purpose is the encompassing goal to help users to actually \emph{make a decision}. In group recommendation scenarios, the latter is usually the case, and understanding human decision-making processes is thus particularly important when designing a GRS. Furthermore, such an understanding is also important when we have to deal with ``high involvement'' application scenarios of recommender systems. A low involvement scenario would be when a system selects and automatically plays music in the background. At the other end of the spectrum there are scenarios where users make deliberate comparisons or interactively select an item in an iterative decision process. Again, group recommendation problems, as envisioned in this paper, more often than not requires a system support when they fall in the high-involvement area of this spectrum, where group members may need multiple interaction rounds before settling on a decision. What adds to the involvement in GRS settings is the potential existence of complex social relationships between group members and other phenomena known from the literature on group dynamics and group decision making~\cite{Eden2020,Forsyth2019}, e.g., when and how group members change their mind.

Regarding the \emph{quality of choices} that are facilitated by a recommender system, the evaluation can also become quite complex for a group recommender system. Individual group members may not only have diverging preferences, but also different beliefs about what represents a desirable decision outcome. Furthermore, a decision might be preferable for some members if it avoids negative effects such as conflicts among group members. In addition, in group recommendation scenarios, there might be an increased desire for \emph{justifiability} of the decision outcome. From the perspective of the decision-making literature~\cite{Eden2020}, we note that it is not only important to justify the decision, but also that the decision-making \emph{process} is perceived as being plausible, fair and transparent. Thus, a GRS should be able to establish these criteria in order to maximize its acceptance.

\paragraph{Supporting Choice Patterns.}
A central element of the work of Jameson et al.~\cite{Jameson2022} is the identification of common \emph{choice patterns}, i.e., typical ways of how users make decisions. Analogously, corresponding mechanisms and strategies that can be implemented to support these choice patterns and the overall decision process are discussed. As an example, consider the pattern of \emph{attribute-based choice}, where users evaluate individual features of each item to make an assessment of its suitability or desirability, either from an individual perspective or from the perspective of the group. To support such a choice pattern, a recommender system may not only provide the relevant information about the features of the items, but may also help the user to focus on a relevant subset of the features, i.e., explain what the decisive criteria might be. In addition, the recommender system might also support side-by-side comparison of different alternatives. In a group recommendation setting, presenting the preferences of individual group members -- per group member or in an aggregated form in a \emph{shared representation} -- may be an effective way of supporting a group's decision process.

Another example of a choice pattern is called \emph{socially based choice}. One form of this social choice pattern in an individual recommendation setting would be the consideration of other users' ratings for certain items. The average rating score of an item on a platform is commonly known to influence individual decisions~\cite{Xitong2018Impact}. In a group recommendation scenario, additional factors come into play. A prime example mentioned above is \emph{conformity}, where group members might adapt their choices according to the examples or expectations of the group. The extent of this behavior can depend on a variety of factors, e.g., if group members know and like each other. An extreme form of such conformity patterns can result in the mostly undesirable phenomenon of \emph{groupthink}, where there is little critical reflection and independent decision-making, which may ultimately lead to poor choices. A related phenomenon in GRS is that of \emph{emotional contagion}, where individuals make choices which they believe may please other group members~\cite{Masthoff2022GRSBeyond}. Potentially one could deal with these phenomena by anonymizing group members. However, such a solution may have detrimental effects on the trust and credibility of the system behavior.

\paragraph{Meta-Level Decision Support.}
In terms of more general choice support techniques, Jameson et al.~for example identify the ``Advise about processing'' strategy. In this strategy, the recommender system gives advice on the meta-level of the decision process. For instance, the GRS might recommend the user to adopt a socially based strategy instead of an attribute-based one for a given group decision situation. A similar idea is proposed in the CAJO model~\cite{ricci2025wideningrolegrouprecommender}, where the role of the first of four proposed agents, the Coach agent, would be to educate group members on how to conduct the group decision-making task. Another technique is named ``Represent the choice situation'', which is particularly relevant in group recommendation scenarios. This strategy aims at establishing a shared representation of the current state of the situation, e.g., by visualizing the individual group members' preferences with respect to item features. Enabling communication among group members can be a key ingredient here as well. Furthermore, when thinking of the envisioned chat-based approaches, summaries of past conversations and expressed preferences may be a suitable mechanism to create a joint understanding of the decision situation. In CAJO~\cite{ricci2025wideningrolegrouprecommender}, the second proposed agent, the Arbiter, is in charge of coordinating the decision-making process and providing crucial information as previously stressed.

\paragraph{Implications for Next-Generation GRS}
These insights suggest that the role of a group recommender system can extend substantially beyond aggregating preferences and identifying suitable items. Particularly in high-involvement decision scenarios, the system may support not only the \emph{outcome} of the decision process, but also the \emph{process} itself. Depending on the situation, this may involve helping group members compare alternatives, establishing a shared understanding of preferences and options, accommodating different choice patterns, or providing guidance at the meta-level on how to proceed with the decision. Importantly, many of these forms of support require an understanding of the evolving decision situation and flexible interaction with the group. As we argue in Section~\ref{sec:research-agenda}, the conversational and reasoning capabilities of modern LLMs create new opportunities to implement such forms of decision support in a more integrated and adaptive manner. Table~\ref{tab:decision-insights} summarizes insights from the decision-making literature and outlines opportunities to support group decision-making with GenAI.

\begin{table*}[t]
\centering
\caption{Selected insights from decision-making research and implications for next-generation GRS.}
\label{tab:decision-insights}
\small

\begin{tabularx}{\textwidth}{p{0.22\textwidth}XX}
\toprule
\textbf{Aspect} & \textbf{Insight from the literature} &
\textbf{Opportunities with LLMs/GenAI} \\
\midrule

Reducing the Choice Set &
Group decisions may require several rounds of discussion and comparison. &
Support iterative deliberation and adapt recommendations to the evolving decision process. \\
\addlinespace

Decision process quality  and Choice patterns&
The quality of the decision process, including justifiability and fairness, matters alongside the outcome. Group members' assessment of items and attributes can in fact be influenced by social factors such as conformity or interpersonal relationships. &
Identify relevant decision criteria and arguments. Summarize evolving preferences, agreements, disagreements, and establish a shared understanding of alternatives and trade-offs. Encourage consideration of diverse perspectives.\\
\addlinespace

Meta-level decision support &
Groups may benefit from guidance on how to structure and conduct their decision process. &
Suggest suitable decision strategies and provide context-sensitive guidance during discussions. \\

\bottomrule
\end{tabularx}
\end{table*}

\subsection{Chat-based Group Recommender Systems}
\label{sec:chatbots-decision-making}
The idea of supporting group decision-making in chat environments has been explored both in recommender systems and related fields.  
In this section, we review selected chat-based group recommender systems.

\paragraph{Existing Chat-based Solutions}
In an early work in this area, Nguyen et al.~\cite{Nguyen2017AChatBased} presented a mobile application for the tourism domain, which aims to support a group of users to choose a point-of-interest (POI) to visit. This work is based on the insights of a preceding observational study~\cite{Delic2016Observing}, which confirmed that group decision-making can be a highly dynamic process, where individual preferences can be constructed and may change during the process. In fact, the general design of the chat-based system is based on insights from the field of group dynamics~\cite{Forsyth2019}. According to such theories, group decision-making is assumed to follow a four-stage process (Orientation–Discussion–Decision–Implementation), and the chat functionality is designed to support the central discussion phase of the process. Besides the possibility of chatting with other group members, the system provides a number of recommendation-related functionalities. For instance, group members can share links to existing POIs as recommendations, to which others can react with a 'like' or 'dislike' or leave a comment. Furthermore, the system can show summaries of group preferences or aggregate elicited ratings to generate a list of recommendations. An initial user study of a prototype system revealed that participants were generally satisfied with the recommendations, their final choice and the system's usability.

Around the same time, Álvarez-Márquez and Ziegler~\cite{alvarez2018negotiation} proposed a related interactive group recommender system with the aim of supporting the entire negotiation process. Unlike in Nguyen et al. ~\cite{Nguyen2017AChatBased}, group members do not propose individual items to others, but they first try to agree on desirable item features, such as the maximum price of a hotel in a tourism use case. They can then exchange their evaluations of the system's recommendations before making a final decision. Meanwhile, group members can freely exchange opinions in a chat area. Several prototypes of the system were developed; the latest was designed as a mobile app. This has a simplified user interface, which was positively assessed in an initial user study. From a system design perspective, while the chat area of the app was assumed to be useful for the negotiation process, it was not deeply integrated with the other system's functionality and screens, and was not the focus of the evaluation.

Building on the chat-based approach by Nguyen et al.~\cite{Nguyen2017AChatBased}, Delic et al.~\cite{Delic2023Charm} more recently proposed the CHARM framework, mentioned earlier. CHARM follows the idea of enabling group discussions in a decision-making setting by incorporating a chat functionality. However, it also introduces an automated chatbot to support the process as a mediator role. In the initial version of the envisioned system, group members can chat and make suggestions freely, e.g. of places of interest to visit in a tourism use case. Other group members can then react to these suggestions by, for example, liking them. The chatbot can be involved by issuing specific chat commands. The chatbot's functions include registering a suggestion, asking for a summary of group member preferences and making a recommendation. From the implementation perspective, the chatbot is meant to be deployed in an existing chat platform, such as WhatsApp or Telegram. However, the CHARM framework has not yet been systematically evaluated.

\paragraph{Implications for Next-Generation GRS}
The discussed works demonstrate that chat-based environments provide a promising setting for supporting group recommendation and decision-making. Existing studies show that group members can successfully discuss preferences and alternatives while recommendation functionality is integrated into the interaction process. However, none of them actively intervenes in the group conversation itself. At the same time, these approaches were largely developed before the recent advances in LLMs and therefore leave substantial room for further development. As summarized in Table~\ref{tab:chat-based-decision-support}, LLMs may, for example, help to interpret ongoing discussions, infer evolving preferences and user intents, summarize the state of the decision process, and provide recommendation support in a natural way. Furthermore, an LLM-supported system could actively intervene in the group discussion, for example, by encouraging less active members to participate or by mediating conflicts, a functionality that is not present in any of the systems discussed above.

\begin{table*}[t]
\centering
\caption{Selected approaches to chat-based group recommender systems and 
opportunities enabled by LLMs and Generative AI.}
\label{tab:chat-based-decision-support}
\small
\begin{tabularx}{\textwidth}{p{0.16\textwidth} X X}
\toprule
\textbf{Work} &
\textbf{Main idea} &
\textbf{Opportunities with LLMs/GenAI} \\
\midrule

Nguyen et al.~\cite{Nguyen2017AChatBased} &
A chat-based GRS supports dynamic group decision-making by allowing  members to discuss and share candidate items, react to them, and obtain  aggregated recommendations. 
& LLMs could continuously interpret the discussion, infer evolving  preferences, and provide recommendation support without requiring  preferences and reactions to be explicitly encoded through dedicated interface elements. \\

\addlinespace
Álvarez-Márquez and Ziegler~\cite{alvarez2018negotiation} &
An interactive GRS supports negotiation by letting group members first agree on desirable item features and subsequently discuss and evaluate system-generated recommendations. &
A tighter integration of conversation and recommendation becomes possible
by automatically extracting relevant criteria, preferences, arguments,
and changes of opinion from the discussion. \\

\addlinespace
Delić et al.~\cite{Delic2023Charm} (CHARM) &
A chatbot embedded in an existing messaging platform mediates group 
decision-making through explicit commands for tasks such as registering 
suggestions, summarizing preferences, and generating recommendations. &
Predefined commands could be replaced by natural-language interaction,
with LLMs recognizing user intents and proactively determining when
and how to support the decision process. \\

\bottomrule
\end{tabularx}
\end{table*}

\subsection{Conversational Agents for Group Decision Support}
\label{sec:conv-agents-gds}

Supporting group decision-making has a long tradition outside the recommender systems field. In Information Systems, research on Group Decision Support Systems (GDSS) dates back to the 1980s~\cite{DeSanctis1987AFoundation,Nunamaker1991Electronig}, with critical reflections in~\cite{PERVAN1998149,Nunamaker1996Lessons}. In parallel, the Computer-Supported Cooperative Work (CSCW) community recognized early on that supporting groups poses different design challenges than supporting individuals~\cite{Baecker1993Readings,Rodden1991Survey}. Early systems such as Colab~\cite{Stefik1987Beyond} and Cognoter~\cite{Foster1986Cognoter} therefore not only served as communication media, but actively structured discussion and decision-making in meetings. Both communities typically take an interdisciplinary socio-technical perspective, and recent works explore the use of LLMs for collective decision-making~\cite{Papachristou2025Leveraging,Enhancing2024Enhancing}.

In a systematic review of 53 studies, drawn primarily from the HCI community, Yeo et al.~\cite{Yeo2026GCA} examine what they call Group Conversational Agents (GCAs), i.e., conversational systems designed to \textit{sense} and \textit{intervene} in group-level interaction, rather than to optimize responses for individual users. The authors organize existing GCAs around four recurring group challenges: (i) orientation (e.g., getting acquainted and establishing norms of
interaction), (ii) conflict (e.g., mediating disagreement), (iii) alignment (e.g., structuring communication and building shared understanding), and (iv) execution (e.g., supporting the ongoing task with information or contributions). Furthermore, they identify three typical roles of such agents: \emph{facilitators or mediators} that regulate how group members interact, \emph{peers} that contribute to the discussion like human participants, and \emph{assistants} that support the group with information from the background. In the following, we review selected works that illustrate these roles and challenges in more depth.

\paragraph{Orientation} Agents addressing \emph{orientation} typically act as social facilitators that help groups initiate interaction. BlahBlahBot~\cite{Shin2021BlahBlahBot} addresses a lack of social familiarity among group members by analyzing their social media content to identify shared interests and suggesting conversation topics. IntroBot~\cite{Shin2023IntroBot} addresses a lack of interaction structure by scaffolding early group interaction through prompts that mark phases of the conversation, such as self-introductions and topic discussion, and further encourages engagement by prompting members who remain inactive.

\paragraph{Alignment} Many works address \emph{alignment}, with the agent acting as a facilitator of task-related discussion. Kim et al.~\cite{Kim2020Bot} tackle coordination problems in group chats: their chatbot monitors participation and proactively encourages less active members to voice their opinion, which was found to increase opinion diversity. Their subsequent DebateBot~\cite{Kim2021ModeratorChatbot} addresses breakdowns in communication processes by structuring deliberative discussions, e.g., by asking discussants to share the reasoning behind their proposals, with positive effects on opinion diversity and the perceived quality of the discussion and decisions. Other systems help groups maintain a shared awareness of the discussion. Tilda~\cite{Zhang2018MakingSense} lets Slack users tag messages, e.g., as questions, ideas, or topic boundaries, and provides summaries that help reconstruct long chat conversations. SolutionChat~\cite{Lee2020SolutionChat} visualizes the state of the discussion for participants and moderators, and supports human moderators by recommending suitable moderation messages. Many functions that had to be hand-coded in these systems, such as message annotation and summarization, can nowadays be automated with LLMs.

\paragraph{Conflict} Agents addressing \emph{conflict} are typically designed as mediators. Govers et al.~\cite{Govers2024Mediation} target conflicts arising from a lack of consensus: their agent mediates online debates by identifying opposing viewpoints, asking probing questions to surface the participants' reasoning, and proposing compromise solutions. Wagner et al.~\cite{Wagner2022Moderation}, in contrast, address a lack of negotiation support in a group scheduling task, comparing public coordination with a strategy in which the agent privately contacts individual members to explore their flexibility.

\paragraph{Execution} Finally, agents addressing \emph{execution} support the group in performing the task itself, either as assistants or as peers. As an assistant, the chatbot by Gürkan and Yan~\cite{Gurkan2023ChatbotCata,Yan2023itdepends} shares additional task-relevant information during team decision-making. Their studies show that the \emph{timing} of such assistance matters: it was most effective, both for information sharing and decision quality, when provided early in the discussion, suggesting that a proactive\footnote{In the literature on conversational recommender systems, the distinction between reactive and proactive roles roughly corresponds to \emph{user-driven} and \emph{system-driven} approaches~\cite{jannach2021crscsur}.} recommendation agent may be most effective when stepping in during early phases. As peers, agents contribute to the discussion like human members. Zheng et al.~\cite{Zheng2023Competent} study an agent that participates as an equal member in a group decision task, sharing its assessments, discussing, and voting, and identify a gap in enabling AI to participate equally in group decision-making. Devil's advocate agents, finally, address diminished critical thinking, i.e., the conformity and groupthink discussed in Section~\ref{sec:human-decision-making}, by challenging either the recommendations of an AI system~\cite{Chiang2024DevilsAdvocate} or the group's emerging majority opinion~\cite{Chiang2024DevilsAdvocate,Lee2025Amplifying,Lee2025Catalysts}.

\medskip
Despite this growing body of work, Yeo et al.~\cite{Yeo2026GCA} observe that current GCAs are predominantly designed as short-term, role-bounded interventions that target a single group challenge, and that they are mostly evaluated in controlled lab settings. They furthermore identify three recurring design tensions: between the \emph{visibility and discretion} of interventions (e.g., addressing a dominating group member publicly or in private), between \emph{proactivity and group autonomy} (i.e., who decides when an intervention is needed), and between \emph{agent authority and group ownership} of the decision. Consistently, a survey of ``polyadic'' chatbots, i.e., conversational agents that support multi-party interactions, reports steadily increasing research interest and finds that such agents typically act as discussion facilitators, counteracting unstructured communication and uneven participation~\cite{kuhal2024polyadic}. While LLMs are only mentioned as future work in~\cite{kuhal2024polyadic}, first LLM-based GCAs are already emerging~\cite{Chiang2024DevilsAdvocate,Addlesee2024,Lee2025Catalysts}; a further recent survey on multi-party conversational agents can be found in~\cite{Sapkota2025MultiParty}.

\paragraph{Implications for Next-Generation GRS} The discussed works demonstrate that conversational agents can actively support group decision-making processes, and they explore the kinds of interventions that are largely missing in existing chat-based GRS. Existing studies show that agents can successfully take on different roles in a group, e.g., as facilitators that encourage participation and structure the discussion, as mediators that help resolve conflicts, or as peers that challenge an emerging consensus. However, as noted above, these agents typically target a single group challenge, are evaluated in short-term lab settings, and are not designed for recommendation scenarios. As summarized in Table~\ref{tab:gca-decision-support}, many of their interventions could be adopted by a group recommendation agent, for example, eliciting preferences from less active members, asking members to justify their preferences, proposing compromise items, or eliciting preferences privately. Furthermore, an LLM-supported recommendation agent could combine these roles and switch between them as the discussion evolves, while balancing the design tensions discussed previously.

\begin{table*}[!t]
\centering
\caption{Overview of selected conversational agents for group decision support from the literature, including challenges, agent roles, and opportunities enabled by Generative AI. Challenges and roles follow the categories of \cite{Yeo2026GCA}; works marked with $^{*}$ are not part of their review and were assigned by us.}
\label{tab:gca-decision-support}
\tiny
\begin{tabularx}{\textwidth}{p{0.14\textwidth} p{0.11\textwidth} X X}
\toprule
\textbf{Work} &
\textbf{Challenge / role} &
\textbf{Main idea} &
\textbf{Opportunities with LLMs/GenAI} \\
\midrule

BlahBlahBot~\cite{Shin2021BlahBlahBot} &
Orientation / social facilitator &
Analyzes participants' social media content to identify shared interests and suggests conversation topics to initiate interaction. &
Shared interests could be inferred from the conversation to help ad-hoc groups start discussing their preferences. \\

\addlinespace
IntroBot~\cite{Shin2023IntroBot} &
Orientation / social facilitator &
Scaffolds early group interaction through prompts that mark conversation phases, and encourages members who remain inactive. &
The early phase of a group decision could be structured, e.g., by eliciting each member's initial preferences through adaptive prompts. \\

\addlinespace
Kim et al.~\cite{Kim2020Bot} &
Alignment / facilitator &
Monitors participation and proactively encourages less active members to voice their opinion, increasing opinion diversity. &
Members whose preferences are not yet known could be detected and asked for their input in a natural, personalized way. \\

\addlinespace
DebateBot~\cite{Kim2021ModeratorChatbot} &
Alignment / facilitator &
Structures deliberative discussions by soliciting opinions from reticent members and asking discussants to share the reasoning behind their proposals. &
Context-sensitive prompts could ask members to justify their preferences, supporting more justifiable group decisions. \\

\addlinespace
Tilda$^{*}$~\cite{Zhang2018MakingSense} &
Alignment / facilitator &
Lets users tag chat messages and provides summaries that help reconstruct long discussions. &
Messages could be annotated automatically, e.g., as proposals or preference statements, and the discussion summarized without manual tagging. \\

\addlinespace
SolutionChat$^{*}$~\cite{Lee2020SolutionChat} &
Alignment / support for human facilitator &
Visualizes the state of the discussion and recommends moderation messages to human moderators. &
The evolving discussion could be interpreted to generate moderation actions, potentially taking on parts of the moderator role. \\

\addlinespace
Govers et al.~\cite{Govers2024Mediation} &
Conflict / mediator &
Mediates online debates by identifying opposing viewpoints, asking probing questions, and proposing compromise solutions. &
Diverging preferences could be made explicit, and compromise items acceptable to all members proposed. \\

\addlinespace
Wagner et al.~\cite{Wagner2022Moderation} &
Conflict / mediator &
Compares public coordination with a private negotiation strategy in which the agent contacts members individually in a group scheduling task. &
Private elicitation could help members state preferences they are reluctant to share publicly, e.g., due to privacy concerns or conformity pressure. \\

\addlinespace
Gürkan and Yan$^{*}$~\cite{Gurkan2023ChatbotCata,Yan2023itdepends} &
Execution / assistant &
A chatbot shares task-relevant information during team decision-making; assistance was most effective when provided early. &
The agent could reason about the state of the discussion to decide when to proactively provide information or recommendations. \\

\addlinespace
Zheng et al.~\cite{Zheng2023Competent} &
Execution / peer &
An agent participates as an equal member in a group decision task, sharing assessments, discussing, and voting. &
The agent could act as an additional group member, contributing item suggestions and opinions alongside human members. \\

\addlinespace
Devil's advocate agents~\cite{Chiang2024DevilsAdvocate,Lee2025Amplifying,Lee2025Catalysts} &
Execution / peer &
Agents challenge AI recommendations or the group's emerging majority opinion to counteract conformity and groupthink. &
Emerging consensus on an item could be challenged, e.g., by pointing out overlooked alternatives or drawbacks. \\

\bottomrule
\end{tabularx}
\end{table*}

\section{Towards Generative AI supported Conversational Group Recommendation} %
\label{sec:research-agenda}
Building on the insights from the different strands of  research reviewed in the previous section, we now turn to our vision of a framework for realizing these forms of group decision support using LLMs and Generative AI. Although several approaches to providing chat-based group decision support have been already proposed in the literature presented in the previous section, these were developed in the pre-LLM era. Thus, they could not benefit from the more recent techniques offered by generative AI, in particular in the area of LLMs. In fact, early approaches were limited by the capabilities of natural language processing (NLP) techniques available at the time. Today, the potential of LLMs, AI assistants like ChatGPT, and Generative AI technology in general, to support and implement traditional recommendation processes have been demonstrated in numerous studies~\cite{Deldjoo2024Review,liu2023chatgpt,hou2023large,Wu2023survey}.

\subsection{Envisioned Framework}
In this section, we outline a vision of next-generation group recommender and decision support systems enabled by Generative AI technology.\footnote{While our current vision mostly builds on LLMs, future systems may also incorporate more general Generative AI models that, for example, support multimodal content.} Our envisioned framework moves beyond the traditional recommendation model in which individual preference profiles are collected, aggregated, and used to generate a recommendation list upon request. In addition to generating recommendations, a GRS implementing the framework can provide various forms of support that may help a group make progress towards a decision, building on the insights discussed in Section~\ref{sec:human-decision-making}.

\subsubsection{Design Rationale and Illustrative Scenario}
We propose a conceptual framework describing \emph{a class of GRS agents} and their decision-support capabilities, rather than prescribing a specific technical implementation. Our vision is to provide a flexible foundation for implementing different conversational group decision-support strategies, as we do not assume the existence of a universally ``best'' decision process or outcome. Accordingly, we envision a configurable approach in which a GRS can be adapted (e.g., through prompting) to support the conversation in the most suitable manner for the given application scenario and group context. Observational studies such as those reported in~\cite{Delic2024Supporting} can help system designers make informed choices, including whether to prioritize the decision outcome or the decision-making process. Ultimately, the selection of appropriate criteria for evaluating the quality of the decision process or its outcomes remains the responsibility of the system designer.

\begin{figure}[h!]
    \centering
    \includegraphics[width=0.45\textwidth]{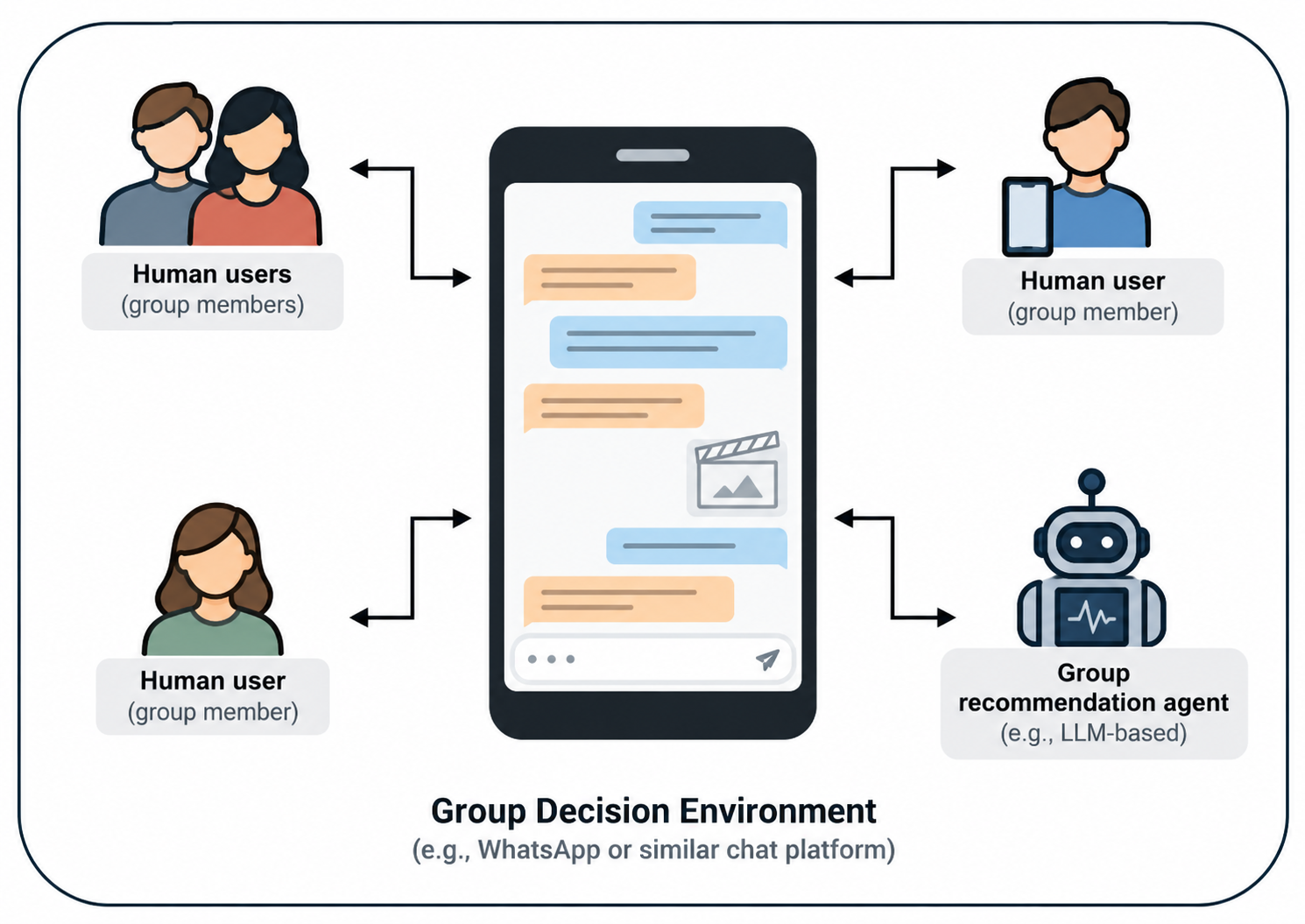}
    \caption{Illustration of an envisioned AI-supported group decision-making scenario in which group members discuss options using a common messaging app, with a group decision-support agent participating in the conversation. The agent's behavior can be configured for the specific use case, ranging from reactive responses to proactive interventions.}
\label{fig:usage-scenario}
\end{figure}

\autoref{fig:usage-scenario} illustrates a typical target application scenario of the envisioned framework. Here, a group of friends is discussing and planning a joint activity, such as attending a cultural event, going to the cinema, or dining out together. In this process, they communicate through their usual instant messaging app. By invitation, a GRS agent, which implements the proposed framework, is added to the chat, similarly to adding the Meta AI chatbot to a WhatsApp group. Depending on the configuration, the GRS can take different roles. 
For example, the system may be configured to make contributions to the conversation mostly in a reactive way, i.e., when being explicitly asked. Alternatively, it can be given a proactive role and, for example, function as a moderator of the discussion. It should be noted that, analogously to \cite{cacm25llmagentplatform}, also multiple recommendation agents, each taking a different role, could facilitate such a group recommendation scenario.

\subsubsection{Decision-Support Capabilities}
\begin{figure}[ht!]
    \centering
    \includegraphics[width=0.99\textwidth]{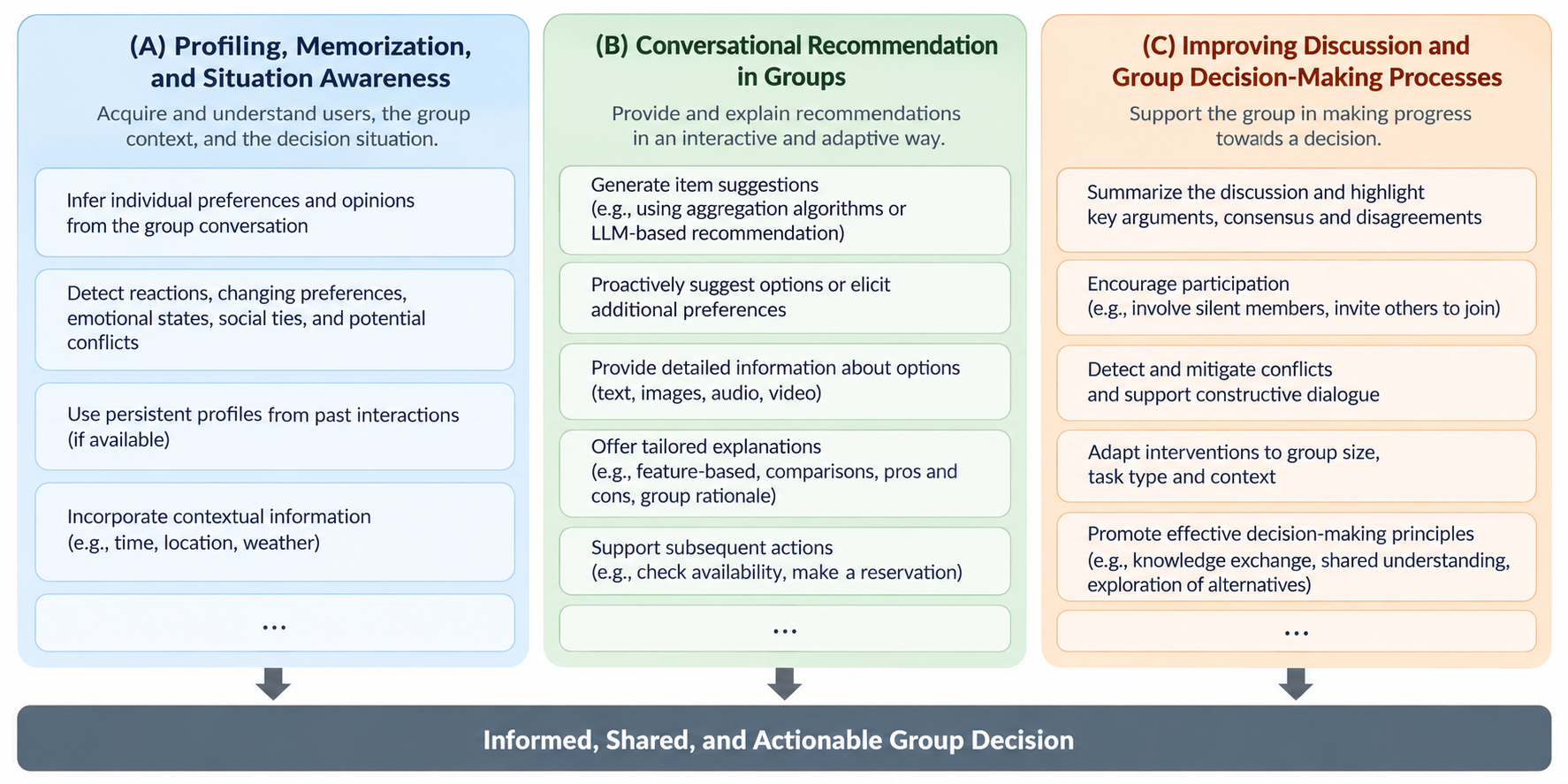}
    \caption{Overview of groups of supported capabilities in the envisioned framework.}
    \label{fig:supported-capabilities}
\end{figure}

The proposed framework encompasses three main areas of functionality: (A) profiling, memorization, and situation awareness; (B) conversational recommendation; and (C) support for discussion and group decision-making. These areas and specific examples of supported functionalities are visualized in Figure~\ref{fig:supported-capabilities}.

\medskip
\noindent\textit{(A) Profiling, Memorization, and Situation Awareness.}
A central task in supporting group decision-making and recommendation is to acquire and understand the needs, preferences, and opinions of individual users in a particular group setting. Existing group recommender systems mainly assume that explicitly expressed user preferences, e.g., through item ratings or preferred item feature values, are known. However, limited research has explored how such preferences could be acquired unobtrusively.

In the envisioned framework, the preferred approach to acquiring user preferences is to leverage the natural language understanding capabilities of LLMs to analyze individual utterances in ongoing group discussions. Such an analysis need not be limited to explicit preference statements but may also consider how group members react to previously expressed preferences. This could allow the GRS to infer information about users' emotional states, social ties between group members, or potential conflicts among them~\cite{Karahodza2025GroupDynamics}. Furthermore, continuously analyzing the discussion could help the system recognize when individual group members change their minds, an aspect that has received limited attention in the existing GRS literature.

If group participants are known from previous interactions, the system may retrieve persistent user profiles from an existing database or an external source. If such information is unavailable, and preferences cannot yet be inferred from the ongoing conversation, the GRS may proactively enter the discussion to acquire additional information. For example, it could ask individual group members to express their preferences or comment on those of others.

To better understand the decision-making situation, the GRS may also acquire additional contextual information relevant to the recommendation process. Such information could include the time of day, the geographic location of group members, or local weather conditions. The GRS could obtain this information by asking corresponding questions in the chat or invoking external tools, such as publicly accessible weather services.

\medskip
\noindent\textit{(B) Supporting Conversational Recommendation in Groups.}
In a traditional GRS, group members first state their preferences, and the system then uses a selected preference aggregation strategy to generate a list of recommendations upon request. The envisioned framework also supports this conventional approach. For example, upon receiving an explicit recommendation request, the GRS may invoke a component implementing a traditional preference aggregation algorithm or rely on the zero-shot or few-shot recommendation capabilities of an LLM~\cite{feng2023largelanguagemodelenhanced}. To present the results, the system may additionally retrieve information about the recommended items from internal or external knowledge sources.

Beyond generating recommendations, in the envisioned framework the GRS supports various other conversational moves~\cite{Tariq2009Improving,jannach2021crscsur} in order to select the most appropriate action for the current situation. For example, instead of passively awaiting a recommendation request, the system may proactively enter the discussion and suggest suitable items~\cite{Rook2020}. Alternatively, it may recognize that additional preference information is needed to generate suitable recommendations and initiate a corresponding elicitation process.

Particularly in group decision-making situations, it is important that group members develop a shared understanding of the available options, the reasons why certain alternatives may be preferable, and the rationale behind the system's suggestions. The capabilities to \emph{inform} and \emph{explain} are therefore central to an LLM-based GRS. Regarding \emph{informing}, the system can provide detailed information about available or recommended items. Such information may be generated using a general-purpose or domain-specific LLM, or retrieved from external sources, such as the Web. The information presented need not be limited to text but may also include multimedia content, such as images, audio, or videos.

Regarding \emph{explanations}, the GRS may dynamically select different explanation approaches depending on their intended purpose~\cite{tintarev2007survey} and the specific group context. Explanations could, for instance, refer to item features or user preferences, or take the form of side-by-side comparisons and lists of pros and cons~\cite{NunesJannachUmuai2017}. The GRS may also invoke external explanation modules implementing specific group explanation strategies. Such group-oriented explanations could clarify why a particular alternative may be beneficial for the group as a whole. Furthermore, LLMs can be used to adapt the content and style of explanations to the needs of different groups.

Ultimately, group decision-making typically aims to reach an agreement on one of the available options, such as selecting a restaurant for dinner. This decision may entail subsequent \emph{actions}, such as making a reservation. In some scenarios, the GRS could also support or carry out these actions. For example, it may check the opening hours of the selected restaurant through an online search or make a reservation using a web form or even a voice call~\cite{GoogleDuplex2018}.

\medskip
\noindent\textit{(C) Improving Discussion and Group Decision-Making Processes.}
Depending on its configuration, in the envisioned framework the GRS may act as a reactive \emph{tool} that supports the decision-making process upon request or take a more active role in the discussion, for example, as an additional virtual group member or a moderator~\cite{Eden2020,Kim2020Bot}. In the latter case, the system would need to determine when to proactively intervene and what to contribute to help the group make progress towards a decision~\cite{Yan2023itdepends}.

A wide range of possible interventions exists, as discussed in Section~\ref{sec:chatbots-decision-making}. For example, the GRS may summarize the current state of the discussion by reviewing suggestions and arguments or highlighting areas of consensus and disagreement. Such summaries may be particularly useful during lengthy or intensive discussions. If the discussion becomes stuck, the system could identify potential leaders and collaborators~\cite{Contreras2021Integrating} or encourage less active group members to express their opinions. Another possibility is to suggest involving additional participants. The system could also detect signs of escalating conflicts and intervene in a moderating role.

When selecting an intervention, the GRS should consider that its effectiveness may depend on factors such as group size and the nature of the decision task~\cite{Kim2020Bot}. Furthermore, the system should follow general \emph{principles} of effective group decision support, as suggested by~\cite{Jameson2022}. For example, it should encourage knowledge exchange among group members, particularly when they differ in their background knowledge or expertise. The system could also help members understand each other's reasoning, develop a shared understanding of the decision-making process, and explore alternatives that accommodate the interests of most or all group members. More generally, an important goal is to facilitate a transparent and inclusive decision-making process in which all members have the opportunity to contribute.

\medskip
The flexibility of the envisioned framework also allows the GRS to support alternative approaches to group decision-making, depending on the specific situation and  needs of the group. To illustrate how the previously discussed capabilities can be combined for this purpose, Table~\ref{tab:principles-and-mechanisms} shows how the framework could support the different group choice patterns identified by~\citet{Jameson2022}.

\begin{table}[h!]
\caption{Illustrative support mechanisms for different group choice patterns~\cite{Jameson2022} within the envisioned framework.}
\label{tab:principles-and-mechanisms}
    \small
\begin{tabular}{p{3cm}p{7.5cm}}
  \toprule
  \textbf{Choice pattern } & \textbf{System Support} \\ \midrule
  
    Attribute-based choice &
    The system can summarize item features, highlight relevant decision criteria, and explain trade-offs between alternatives. \\
    \addlinespace
    
    Socially-based choice &
    The system can present arguments for and against alternatives raised by others and help group members consider each other's expectations and opinions. \\
    \addlinespace
    
    Consequence-based choice &
    The system can help the group explore alternative courses of action and assess their potential consequences (e.g., regarding sustainability). \\
    \addlinespace
    
    Policy-based choice &
    The system can help the group select an appropriate decision-making procedure based on the context and guide its implementation. \\
    \addlinespace
    
    Experience-based choice &
    The system can recall similar past decisions and help the group assess whether previously used decision strategies are suitable for the current situation. \\
    \addlinespace
    
    Trial-and-error-based choice &
    The system can support the exploration of alternatives, potentially through simulations, and help the group evaluate their implications before making a final choice. \\

  \bottomrule
\end{tabular}

\end{table}

\subsubsection{Route to Implementation}
\label{subsubsec:route-to-implementation}
\begin{figure}[ht!]
    \centering
    \includegraphics[width=0.8\textwidth]{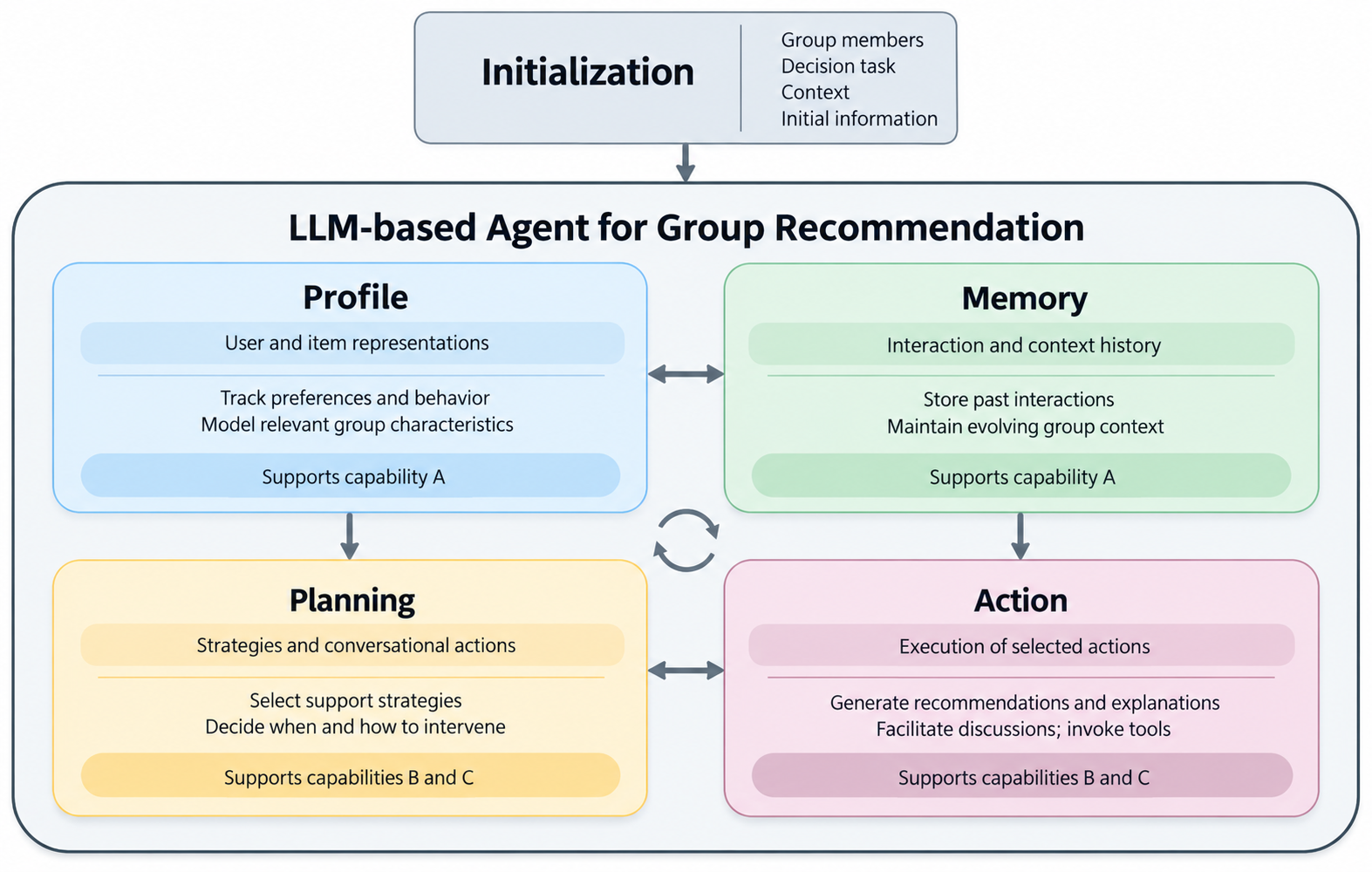}
    \caption{Illustrative agent-based architecture for implementing the envisioned GRS capabilities, following~\cite{peng2025surveyllmpoweredagentsrecommender}.}
    \label{fig:grouprs-agent}
\end{figure}

One viable route to implementing the envisioned decision-support capabilities is to leverage recent approaches for building LLM-based agents~\cite{Wang2024ASurvey,huang2024understandingplanningllmagents}. Such agents can go beyond reactive responses to user prompts by planning future actions, generating explanations, and invoking external tools to accomplish specific tasks. \citet{Wang2024ASurvey} propose a general architecture for LLM-based autonomous agents consisting of four key components: Profile, Memory, Planning, and Action (cf.~Figure~\ref{fig:grouprs-agent}). This architecture was subsequently adopted by~\citet{peng2025surveyllmpoweredagentsrecommender} in their review of LLM-based agents for recommender systems. We consider this architecture a possible technical foundation for implementing the decision-support capabilities outlined above, rather than a necessary component of our conceptual framework.\footnote{Our conceptual vision for future group recommender systems is closely tied to the capabilities of modern LLMs, which have made several forms of decision support realistically feasible that were difficult to implement in the pre-LLM era~\cite{jannach2022grandchallenge}. Nevertheless, certain elements of our framework could also be implemented using alternative approaches, such as traditional planning and recommendation algorithms or neuro-symbolic methods.}

The four components can support different aspects of the envisioned framework. The \emph{Profile Module} maintains representations of users and items. User profiles may capture observed preferences, past behavioral patterns, and potentially social or personal characteristics, as proposed, for example, in~\cite{Karahodza2025GroupDynamics}. The \emph{Memory Module} stores information from previous interactions, potentially including contextual information and users' emotional responses~\cite{peng2025surveyllmpoweredagentsrecommender}. Together, these components provide a basis for the profiling, memorization, and situation awareness capabilities discussed above under \emph{(A) Profiling, Memorization, and Situation Awareness}.

The \emph{Planning Module} determines suitable strategies and corresponding action plans, potentially balancing short-term and long-term goals. Planning may also operate at the level of individual conversational actions, determining when and how the system should intervene in a group discussion. The \emph{Action Module}, finally, executes the selected actions, drawing on information provided by the other components. Beyond generating recommendations, it may provide explanations, contribute to group discussions, or invoke external tools to retrieve information and perform subsequent tasks. Together, Planning and Action can thus support both conversational recommendation and the broader group decision-making processes described in the capabilities \emph{(B) Conversational
Recommendation} and \emph{(C) Support for Discussion and Group Decision-Making} from above.

\subsection{Assumptions, Scope, and Limitations}

The discussions in the preceding sections rest on several implicit and explicit assumptions. We summarize them here to clarify our vision. A more detailed examination of the associated implementation challenges and limitations is provided in Section~\ref{sec:challenges}.

\begin{itemize}
\item \emph{Usage Context:} We assume an instant-messaging, chat-based setting (e.g., WhatsApp-like environments) as a natural locus for group deliberation and decision support, where a bot-like agent can participate in the conversation. This assumption is motivated by our aim to situate group recommendations in a ``convenient'' software environment and to help explain the limited practical adoption of prior GRS approaches that are detached from such everyday contexts. We acknowledge that this approach may be less suited in certain scenarios like communal background music selection (e.g., MusicFX~\cite{McCarthy1998MusicFX}), where other modalities or interfaces might be preferable.
\item \emph{Normative Stance:} We do not prescribe a single ``good'' decision process or outcome, and we do not introduce a central evaluation component that operationalizes such norms. Instead, we acknowledge varying notions of success and make agent behavior configurable to different decision contexts and values. Where appropriate, the agent may also deliver a traditional aggregated ranked list within the chat environment.
\item \emph{Class-of-systems Approach:} Our vision calls for a configurable framework (``a class of systems'') rather than a single, one-size-fits-all solution. The group recommendation agent can be tuned to be more proactive, e.g., by encouraging silent members, or more reactive, depending on the use case.
\item \emph{Use of LLMs:} We focus on LLMs because they are a natural choice for text-centric tasks, e.g., intent recognition, summarization, explanation, and enable agentic capabilities like moderating the discussion, identifying unheard members or de-escalating conflicts. At the same time, our vision is compatible with hybrid or non-LLM implementations for specific components, e.g., logic-based or neurosymbolic reasoning for deciding when or how the agent intervenes.
\item \emph{Relation to Adjacent Fields:} The envisioned system intentionally overlaps with group decision-support systems, e.g., facilitating group discussion through summaries or prompts, but remains anchored in recommendation by always offering a recommendation mechanism, including traditional preference aggregation and information retrieval components, if desired. The system also connects to conversational recommenders, sharing similarities like problems of intent detection, yet being distinct, e.g., in terms of providing consensus mechanisms.
\end{itemize}

Generally, we emphasize that this essay offers a forward-looking perspective, articulating a vision grounded in existing lines of research and acknowledging uncertainty and alternative trajectories. Accordingly, we aim to present this vision as a practical starting point for near-term exploration, rather than a universal prescription.

\section{Challenges}
\label{sec:challenges}
Next, we discuss different types of challenges that must be tackled when implementing and evaluating the envisioned next generation of group recommender systems.

\paragraph{Technical Challenges}
Building software applications that are based on LLMs or integrate LLMs have a number of known challenges, including in particular the problem of hallucinations, the predictability of the outcomes, the lack of reproducibility due to the non-deterministic behavior of LLMs, potential response latency, or high computational costs. Using LLMs to build autonomous agents, as sketched in here for group decision-making scenarios, is a rather new approach, which may come with its own particular challenges. For instance, such agents are typically designed to interact with individual users, which makes intent detection relatively straightforward. However, in group settings, identifying intents, preferences, and opinions is more complex, as it involves analyzing both the interactions that occur between group members and those that happen between individual members and the agent. In \cite{Wang2024ASurvey}, Wang et al.~provide a recent survey on agentic LLMs, and they outline various open issues.

A key challenge is that the agent needs to be trained on sufficient data to perform the required ``role-playing'' functionality. In the case of group recommendation support, for example, there might not be enough information in the corpus on which the LLM was trained to teach the agent how to act as a group moderator. A particular additional problem, as noted in~\cite{fischer2023reflectivelinguisticprogrammingrlp}, is current LLMs lacking means to mimic self-awareness in conversational settings and are therefore unable to model human behavior effectively.
For example, they may lack a clear agenda or underlying motivation.

Another technical challenge arises from the complexity of LLM-based agentic systems, which commonly involve a number of separate modules, e.g., for memory or planning. Thus, the entire system is based on a complex prompt architecture that is needed to enable a proper communication and coordination between these modules. The design of such a prompt framework is seen as particularly demanding because of a potentially limited robustness of different LLMs, where small changes in the input prompts may have major consequences. The development of the prompt framework therefore may require extensive manual engineering and an inefficient trial-and-error development process~\cite{Wang2024ASurvey}.

In terms of the reasoning and planning capabilities of the system, it is known that LLMs have major limitations. While current LLMs may be able to simulate reasoning to a certain extent, they have difficulties in creating consistent multi-step plans, which provable lead to the desired goal. While recent approaches have proposed \emph{learning-based} ways of constructing plans~\cite{Shi2024Large}, some researchers trust the possibility to integrating traditional, \emph{symbolic} and search-based planning algorithms into the overall architecture~\cite{huang2024understandingplanningllmagents}. One advantage of using well-understood, sound and complete planning algorithms is that they reliably return feasible plans. However, precisely describing the planning problem at the required level of detail in a declarative language may require extensive human effort. In this context, neurosymbolic approaches have received increasing attention in recent years, as they aim to combine the strengths of machine learning with symbolic reasoning capabilities~\cite{hitzler2022neurosymbolic,Bhuyan2024}. We emphasize that our envisioned framework is not restricted to using LLMs for all its components. Instead, individual modules may also leverage alternative methods, including traditional planning and reasoning techniques as well as neurosymbolic approaches, depending on the specific requirements of the task.

Finally, in addition to questions related to successfully implementing the agent's behavior, questions regarding the user interface of the group recommender system should also be considered. In our scenario, we assume that group members converse via an instant messaging app with which users are familiar. However, even in a relatively simple chat-based interaction model, various small design choices can impact the user experience of the system, see~\cite{Zhang2018MakingSense,Iovine2020Conversational} for related studies.

\paragraph{Understanding Group-Decision Making in Online Environments}
In order to optimally support a group of users in their decision-making process, it is important to understand the dynamics and mechanics of such processes. Various phenomena related to \emph{group dynamics} have been studied for decades in particular in the field of social psychology~\cite{Forsyth2019}. Relevant phenomena of group dynamics, as discussed in~\cite{Forsyth2019}, for example relate to the hierarchical structure and communication patterns in groups, the different roles the members of a group can take, how group members may influence each other, how conflicts can emerge and how they can be addressed, and in particular how groups make decisions and what pitfalls there might be, e.g., in terms of \emph{groupthink}. The presence of a recommendation agent as part of the group may introduce additional dynamics and open new questions as discussed in Section~\ref{sec:chatbots-decision-making} and in~\cite{Jameson2022}. Besides understanding group dynamics in general, previous research in the social sciences has investigated specific questions related to group negotiation and decision-making~\cite{kilgour2019handbook}. \citet{Eden2020}, for example, discusses behavioral considerations regarding group support mechanisms; \citet{Martinovski2021} reviews the role of emotion in group decisions.
Furthermore, there is a comparably rich literature in the Information Systems field on the topic of Group Decision Support Systems (GDDS)\footnote{The area of Computer Supported Collaborative Work also addresses related problems.}, with roots dating back to the 1980s.

Overall, to build an effective group recommender system it thus seems necessary to adopt an interdisciplinary approach~\cite{Keestra_Menken_2016} to leverage the existing knowledge and insights about group dynamics from the social sciences. In that context, it is important that experts from other disciplines, e.g., from social psychology, are actively involved throughout the entire process of designing, implementing and evaluating the resulting system~\cite{Aboelela2007Defining}.

\paragraph{Evaluation}
In the research literature, we often find that offline experiments are used, as it is commonly done for single-user recommendation, focusing on the evaluation of different recommendation algorithms.\footnote{Reference works like~\cite{Felfernig2018GRSIntro} or~\cite{Masthoff2022GRSBeyond} are strongly focusing on such evaluations as well.} However, as outlined in this essay, the presentation of a list of recommendations based on the aggregated preferences of the group members only covers one part of the functionalities of our envisioned group recommendation agents, which support the group's decision-making process in various ways. Furthermore, the question remains to what extent general system evaluation measures, like precision and recall, may trustworthy inform us of the effectiveness of a group recommender system.

When assessing the quality of a group's decision or choice, in~\citet{Jameson2022} four desirable criteria are proposed: that the ``outcome is good'', that the decision process is efficient, that the process is not unpleasant, and that the decision can be justified. One main challenge for the first criterion is that it may not always be clear at decision time what constitutes a good outcome, as group members may have different preferences and beliefs. In~\citet{Masthoff2022GRSBeyond} the authors argue that the group member's \emph{satisfaction} with the decision is an important factor. Yet, it is clear that the decision that maximizes satisfaction in the group may in the end not be the one with the best outcome, knowing that psychological factors and group dynamics may influence user satisfaction.

Overall, these considerations call for multi-faceted human-centric evaluation approaches, given that aspects like user satisfaction cannot be directly measured in offline experiments. Since the group decision-making process is a conversational activity, existing approaches for the evaluation of conversational recommender systems (CRS) can be applied. A review of evaluation dimensions and corresponding measurement approaches for CRS can be found in~\cite{jannach2022evaluatingcrs} and~\cite{jannach2021crscsur}. Following the discussion above, the evaluation of CRS is not limited to recommendation quality, but extends to choice satisfaction and to the efficiency and quality of the decision-making process as well. Given the challenges of offline evaluations, in~\citet{Jin2024crsque} the authors propose an extension of the widely used ResQue framework for human-centric evaluation for the case of conversational recommendation.

In addition to the challenges of assessing the quality of the decision process and outcome, the inclusion of LLM-based components in the overall architecture comes with further challenges, e.g., because of the stochastic nature of such models. More research on establishing appropriate evaluation methods for LLM-based applications is still required, in particular when LLM agents are involved. Recent surveys on the evaluation of LLM agents, also with a focus on multi-turn conversations, can be found in~\cite{yehudai2025surveyevaluationllmbasedagents,guan2025evaluatingllmbasedagentsmultiturn}. However, these surveys again demonstrate a focus on offline experiments.

For a comprehensive assessment of the quality of a group recommendation agent, studies with users should probably be the method of choice in most future research efforts, as only such studies can inform us reliably about main quality factors such as choice satisfaction or the perceived fairness and transparency of the decision process. Simulations may furthermore represent a viable complementary approach in various situations. In particular, approaches seem promising where the behaviors of individual group members are simulated with the help of LLM-based agents, and where each LLM-based agent has their own preferences, beliefs, social behavior, personality and negotiation strategy in the group conversation.
Notably, the use of \emph{LLMs as user simulators} in conversational recommender systems has gained growing attention in recent years~\cite{Balog2023UserSimulation, Zhu20225ALLMbased, Zhang2025EvalAgent, Nolwenn2026UserSimCRS}, see also~\cite{Petruzzelli2026Generative} for a recent survey. These approaches however are still in their infancy and may exhibit certain limitations or implement assumptions that lead to unrealistic behavior~\cite{kim-etal-2025-stop}.\footnote{A survey on the use of LLMs for general user simulation and a discussion of challenges can be found in~\cite{ni-etal-2026-survey}.}

Furthermore, in addition to controlled experiments and simulations, other forms of research, e.g., based on \emph{observational} methods, can be applied to understand the effects of agent-support group recommendations, see, e.g.,~\cite{Gurkan2023ChatbotCata,Delic2016Observing}. Overall, given the rapid development of the capabilities of LLMs in recent years, our research methodologies must continuously evolve to enable comprehensive assessment of emerging LLM-based solutions and the group recommendation approaches envisioned in this essay.

\section{Summary}
In this \emph{perspectives} paper, we advocate for a re-orientation of the research in group recommender systems for two main reasons. On one hand, even after several years of research on group recommender systems, the commercial and industrial adoption of such systems remains limited. This observation should urge us to question whether the envisioned use cases and decision-making processes that are assumed in the literature are realistic. On the other hand, the rapid development of LLMs and in particular the emerging capabilities of LLM-based agents open entirely new opportunities to support group-decision processes beyond the automated generation of recommendation lists.

In our essay, we have outlined one possible future direction for LLM-based group recommender systems, where a group of users should be supported in a very natural way through the entire decision-making process. The main focus of the envisioned system was to support the decision-making process within a chat-based environment. These ideas could potentially be extended to other platforms in the future, including online group decision-making video calls, in which the AI agent that makes recommendations would participate in the form of an avatar.

Ultimately, we hope that our vision will help stimulate further research in this important area. The integration of modern AI technologies not only raises new technical questions, but also places greater emphasis on human–computer interaction and related aspects. As such, this line of work offers an opportunity for interdisciplinary research and may help attract a broader range of scholars, compared to what has been observed in the past.

\bibliographystyle{ACM-Reference-Format}

\begin{thebibliography}{101}


\ifx \showCODEN    \undefined \def \showCODEN     #1{\unskip}     \fi
\ifx \showDOI      \undefined \def \showDOI       #1{#1}\fi
\ifx \showISBNx    \undefined \def \showISBNx     #1{\unskip}     \fi
\ifx \showISBNxiii \undefined \def \showISBNxiii  #1{\unskip}     \fi
\ifx \showISSN     \undefined \def \showISSN      #1{\unskip}     \fi
\ifx \showLCCN     \undefined \def \showLCCN      #1{\unskip}     \fi
\ifx \shownote     \undefined \def \shownote      #1{#1}          \fi
\ifx \showarticletitle \undefined \def \showarticletitle #1{#1}   \fi
\ifx \showURL      \undefined \def \showURL       {\relax}        \fi
\providecommand\bibfield[2]{#2}
\providecommand\bibinfo[2]{#2}
\providecommand\natexlab[1]{#1}
\providecommand\showeprint[2][]{arXiv:#2}

\bibitem[Aboelela et~al\mbox{.}(2007)]%
        {Aboelela2007Defining}
\bibfield{author}{\bibinfo{person}{Sally~W. Aboelela}, \bibinfo{person}{Elaine
  Larson}, \bibinfo{person}{Suzanne Bakken}, \bibinfo{person}{Olveen
  Carrasquillo}, \bibinfo{person}{Allan Formicola}, \bibinfo{person}{Sherry~A.
  Glied}, \bibinfo{person}{Janet Haas}, {and} \bibinfo{person}{Kristine~M.
  Gebbie}.} \bibinfo{year}{2007}\natexlab{}.
\newblock \showarticletitle{Defining Interdisciplinary Research: Conclusions
  from a Critical Review of the Literature}.
\newblock \bibinfo{journal}{\emph{Health Services Research}}
  \bibinfo{volume}{42} (\bibinfo{year}{2007}), \bibinfo{pages}{329--346}.
\newblock
\urldef\tempurl%
\url{https://doi.org/10.1111/j.1475-6773.2006.00621.x}
\showDOI{\tempurl}


\bibitem[Addlesee et~al\mbox{.}(2024)]%
        {Addlesee2024}
\bibfield{author}{\bibinfo{person}{Angus Addlesee}, \bibinfo{person}{Neeraj
  Cherakara}, \bibinfo{person}{Nivan Nelson}, \bibinfo{person}{Daniel
  Hern{\'a}ndez~Garc{\'\i}a}, \bibinfo{person}{Nancie Gunson},
  \bibinfo{person}{Weronika Siei{\'n}ska}, \bibinfo{person}{Marta Romeo},
  \bibinfo{person}{Christian Dondrup}, {and} \bibinfo{person}{Oliver Lemon}.}
  \bibinfo{year}{2024}\natexlab{}.
\newblock \showarticletitle{A multi-party conversational social robot using
  llms}. In \bibinfo{booktitle}{\emph{Companion of the 2024 ACM/IEEE
  International Conference on Human-Robot Interaction}}.
  \bibinfo{pages}{1273--1275}.
\newblock


\bibitem[Baecker(1993)]%
        {Baecker1993Readings}
\bibfield{author}{\bibinfo{person}{Ronald~M Baecker}.}
  \bibinfo{year}{1993}\natexlab{}.
\newblock \bibinfo{booktitle}{\emph{Readings in groupware and
  computer-supported cooperative work: Assisting human-human collaboration}}.
\newblock \bibinfo{publisher}{Morgan Kaufmann}.
\newblock


\bibitem[Balog and Zhai(2023)]%
        {Balog2023UserSimulation}
\bibfield{author}{\bibinfo{person}{Krisztian Balog} {and}
  \bibinfo{person}{ChengXiang Zhai}.} \bibinfo{year}{2023}\natexlab{}.
\newblock \showarticletitle{User Simulation for Evaluating Information Access
  Systems}. In \bibinfo{booktitle}{\emph{Proceedings of the Annual
  International ACM SIGIR Conference on Research and Development in Information
  Retrieval in the Asia Pacific Region}} \emph{(\bibinfo{series}{SIGIR-AP
  '23})}. \bibinfo{pages}{302–305}.
\newblock
\urldef\tempurl%
\url{https://doi.org/10.1145/3624918.3629549}
\showDOI{\tempurl}


\bibitem[Bernard and Balog(2026)]%
        {Nolwenn2026UserSimCRS}
\bibfield{author}{\bibinfo{person}{Nolwenn Bernard} {and}
  \bibinfo{person}{Krisztian Balog}.} \bibinfo{year}{2026}\natexlab{}.
\newblock \showarticletitle{UserSimCRS v2: Simulation-Based Evaluation for
  Conversational Recommender Systems}. In \bibinfo{booktitle}{\emph{Advances in
  Information Retrieval}}. \bibinfo{publisher}{Springer Nature Switzerland},
  \bibinfo{pages}{496--510}.
\newblock
\showISBNx{978-3-032-21321-1}


\bibitem[Bhuyan et~al\mbox{.}(2024)]%
        {Bhuyan2024}
\bibfield{author}{\bibinfo{person}{Bikram~Pratim Bhuyan}, \bibinfo{person}{Amar
  Ramdane-Cherif}, \bibinfo{person}{Ravi Tomar}, {and} \bibinfo{person}{T.~P.
  Singh}.} \bibinfo{year}{2024}\natexlab{}.
\newblock \showarticletitle{Neuro-symbolic artificial intelligence: a survey}.
\newblock \bibinfo{journal}{\emph{Neural Computing and Applications}}
  \bibinfo{volume}{36}, \bibinfo{number}{21} (\bibinfo{year}{2024}),
  \bibinfo{pages}{12809--12844}.
\newblock
\showISSN{1433-3058}
\urldef\tempurl%
\url{https://doi.org/10.1007/s00521-024-09960-z}
\showDOI{\tempurl}


\bibitem[Chen et~al\mbox{.}(2013)]%
        {Chen2013Human}
\bibfield{author}{\bibinfo{person}{Li Chen}, \bibinfo{person}{Marco de Gemmis},
  \bibinfo{person}{Alexander Felfernig}, \bibinfo{person}{Pasquale Lops},
  \bibinfo{person}{Francesco Ricci}, {and} \bibinfo{person}{Giovanni
  Semeraro}.} \bibinfo{year}{2013}\natexlab{}.
\newblock \showarticletitle{Human Decision Making and Recommender Systems}.
\newblock \bibinfo{journal}{\emph{ACM Trans. Interact. Intell. Syst.}}
  \bibinfo{volume}{3}, \bibinfo{number}{3} (\bibinfo{year}{2013}).
\newblock
\urldef\tempurl%
\url{https://doi.org/10.1145/2533670.2533675}
\showDOI{\tempurl}


\bibitem[Chiang et~al\mbox{.}(2024a)]%
        {Enhancing2024Enhancing}
\bibfield{author}{\bibinfo{person}{Chun-Wei Chiang}, \bibinfo{person}{Zhuoran
  Lu}, \bibinfo{person}{Zhuoyan Li}, {and} \bibinfo{person}{Ming Yin}.}
  \bibinfo{year}{2024}\natexlab{a}.
\newblock \showarticletitle{Enhancing AI-Assisted Group Decision Making through
  LLM-Powered Devil's Advocate}. In \bibinfo{booktitle}{\emph{Proceedings of
  the 29th International Conference on Intelligent User Interfaces}}
  (Greenville, SC, USA) \emph{(\bibinfo{series}{IUI '24})}.
  \bibinfo{publisher}{Association for Computing Machinery},
  \bibinfo{address}{New York, NY, USA}, \bibinfo{pages}{103–119}.
\newblock
\urldef\tempurl%
\url{https://doi.org/10.1145/3640543.3645199}
\showDOI{\tempurl}


\bibitem[Chiang et~al\mbox{.}(2024b)]%
        {Chiang2024DevilsAdvocate}
\bibfield{author}{\bibinfo{person}{Chun-Wei Chiang}, \bibinfo{person}{Zhuoran
  Lu}, \bibinfo{person}{Zhuoyan Li}, {and} \bibinfo{person}{Ming Yin}.}
  \bibinfo{year}{2024}\natexlab{b}.
\newblock \showarticletitle{Enhancing ai-assisted group decision making through
  llm-powered devil's advocate}. In \bibinfo{booktitle}{\emph{Proceedings of
  the 29th International Conference on Intelligent User Interfaces}}.
  \bibinfo{pages}{103--119}.
\newblock


\bibitem[Church and de~Oliveira(2013)]%
        {Church2013}
\bibfield{author}{\bibinfo{person}{Karen Church} {and} \bibinfo{person}{Rodrigo
  de Oliveira}.} \bibinfo{year}{2013}\natexlab{}.
\newblock \showarticletitle{What's up with Whatsapp? Comparing mobile instant
  messaging behaviors with traditional SMS}. In
  \bibinfo{booktitle}{\emph{Proceedings of the 15th International Conference on
  Human-Computer Interaction with Mobile Devices and Services}}
  \emph{(\bibinfo{series}{MobileHCI '13})}. \bibinfo{pages}{352–361}.
\newblock
\urldef\tempurl%
\url{https://doi.org/10.1145/2493190.2493225}
\showDOI{\tempurl}


\bibitem[Contreras et~al\mbox{.}(2021)]%
        {Contreras2021Integrating}
\bibfield{author}{\bibinfo{person}{David Contreras}, \bibinfo{person}{Maria
  Salam\'{o}}, {and} \bibinfo{person}{Ludovico Boratto}.}
  \bibinfo{year}{2021}\natexlab{}.
\newblock \showarticletitle{Integrating Collaboration and Leadership in
  Conversational Group Recommender Systems}.
\newblock \bibinfo{journal}{\emph{ACM Trans. Inf. Syst.}} \bibinfo{volume}{39},
  \bibinfo{number}{4} (\bibinfo{date}{Aug.} \bibinfo{year}{2021}).
\newblock
\urldef\tempurl%
\url{https://doi.org/10.1145/3462759}
\showDOI{\tempurl}


\bibitem[Dara et~al\mbox{.}(2019)]%
        {Dara2019Survey}
\bibfield{author}{\bibinfo{person}{Sriharsha Dara},
  \bibinfo{person}{C.~Ravindranath Chowdary}, {and} \bibinfo{person}{Chintoo
  Kumar}.} \bibinfo{year}{2019}\natexlab{}.
\newblock \showarticletitle{A survey on group recommender systems}.
\newblock \bibinfo{journal}{\emph{Journal of Intelligent Information Systems}}
  \bibinfo{volume}{54} (\bibinfo{year}{2019}), \bibinfo{pages}{271 -- 295}.
\newblock
\urldef\tempurl%
\url{https://api.semanticscholar.org/CorpusID:57427621}
\showURL{%
\tempurl}


\bibitem[Deldjoo et~al\mbox{.}(2024)]%
        {Deldjoo2024Review}
\bibfield{author}{\bibinfo{person}{Yashar Deldjoo}, \bibinfo{person}{Zhankui
  He}, \bibinfo{person}{Julian McAuley}, \bibinfo{person}{Anton Korikov},
  \bibinfo{person}{Scott Sanner}, \bibinfo{person}{Arnau Ramisa},
  \bibinfo{person}{Ren\'{e} Vidal}, \bibinfo{person}{Maheswaran Sathiamoorthy},
  \bibinfo{person}{Atoosa Kasirzadeh}, {and} \bibinfo{person}{Silvia Milano}.}
  \bibinfo{year}{2024}\natexlab{}.
\newblock \showarticletitle{A Review of Modern Recommender Systems Using
  Generative Models (Gen-RecSys)}. In \bibinfo{booktitle}{\emph{Proceedings of
  the 30th ACM SIGKDD Conference on Knowledge Discovery and Data Mining}}
  (Barcelona, Spain) \emph{(\bibinfo{series}{KDD '24})}.
  \bibinfo{pages}{6448–6458}.
\newblock
\urldef\tempurl%
\url{https://doi.org/10.1145/3637528.3671474}
\showDOI{\tempurl}


\bibitem[Delic et~al\mbox{.}(2023)]%
        {Delic2023Charm}
\bibfield{author}{\bibinfo{person}{Amra Delic}, \bibinfo{person}{Hanif
  Emamgholizadeh}, {and} \bibinfo{person}{Francesco Ricci}.}
  \bibinfo{year}{2023}\natexlab{}.
\newblock \showarticletitle{CHARM: A Group Recommender ChatBot}. In
  \bibinfo{booktitle}{\emph{Adjunct Proceedings of the 31st ACM Conference on
  User Modeling, Adaptation and Personalization}} \emph{(\bibinfo{series}{UMAP
  '23 Adjunct})}. \bibinfo{pages}{275–282}.
\newblock
\urldef\tempurl%
\url{https://doi.org/10.1145/3563359.3597388}
\showDOI{\tempurl}


\bibitem[Delic et~al\mbox{.}(2024)]%
        {DelicE0M24}
\bibfield{author}{\bibinfo{person}{Amra Delic}, \bibinfo{person}{Hanif
  Emamgholizadeh}, \bibinfo{person}{Francesco Ricci}, {and}
  \bibinfo{person}{Judith Masthoff}.} \bibinfo{year}{2024}\natexlab{}.
\newblock \showarticletitle{Supporting Group Decision-Making: Insights from a
  Focus Group Study}. In \bibinfo{booktitle}{\emph{Proceedings of the 32nd
  {ACM} Conference on User Modeling, Adaptation and Personalization, {UMAP}
  2024}}. \bibinfo{pages}{301--306}.
\newblock
\urldef\tempurl%
\url{https://doi.org/10.1145/3627043.3659538}
\showDOI{\tempurl}


\bibitem[Deli\'{c} et~al\mbox{.}(2024)]%
        {Delic2024Supporting}
\bibfield{author}{\bibinfo{person}{Amra Deli\'{c}}, \bibinfo{person}{Hanif
  Emamgholizadeh}, \bibinfo{person}{Francesco Ricci}, {and}
  \bibinfo{person}{Judith Masthoff}.} \bibinfo{year}{2024}\natexlab{}.
\newblock \showarticletitle{Supporting Group Decision-Making: Insights from a
  Focus Group Study}. In \bibinfo{booktitle}{\emph{Proceedings of the 32nd ACM
  Conference on User Modeling, Adaptation and Personalization}}
  \emph{(\bibinfo{series}{UMAP '24})}. \bibinfo{pages}{301–306}.
\newblock
\urldef\tempurl%
\url{https://doi.org/10.1145/3627043.3659538}
\showDOI{\tempurl}


\bibitem[Delic et~al\mbox{.}(2018)]%
        {Delic2018Observational}
\bibfield{author}{\bibinfo{person}{Amra Delic}, \bibinfo{person}{Julia
  Neidhardt}, \bibinfo{person}{Thuy~Ngoc Nguyen}, {and}
  \bibinfo{person}{Francesco Ricci}.} \bibinfo{year}{2018}\natexlab{}.
\newblock \showarticletitle{An observational user study for group recommender
  systems in the tourism domain}.
\newblock \bibinfo{journal}{\emph{J. Inf. Technol. Tour.}}
  \bibinfo{volume}{19}, \bibinfo{number}{1-4} (\bibinfo{year}{2018}),
  \bibinfo{pages}{87--116}.
\newblock
\urldef\tempurl%
\url{https://doi.org/10.1007/S40558-018-0106-Y}
\showDOI{\tempurl}


\bibitem[Delic et~al\mbox{.}(2016)]%
        {Delic2016Observing}
\bibfield{author}{\bibinfo{person}{Amra Delic}, \bibinfo{person}{Julia
  Neidhardt}, \bibinfo{person}{Thuy~Ngoc Nguyen}, \bibinfo{person}{Francesco
  Ricci}, \bibinfo{person}{Laurens Rook}, \bibinfo{person}{Hannes Werthner},
  {and} \bibinfo{person}{Markus Zanker}.} \bibinfo{year}{2016}\natexlab{}.
\newblock \showarticletitle{Observing Group Decision Making Processes}. In
  \bibinfo{booktitle}{\emph{Proceedings of the 10th ACM Conference on
  Recommender Systems}} \emph{(\bibinfo{series}{RecSys '16})}.
  \bibinfo{pages}{147–150}.
\newblock
\urldef\tempurl%
\url{https://doi.org/10.1145/2959100.2959168}
\showDOI{\tempurl}


\bibitem[DeSanctis and Gallupe(1987)]%
        {DeSanctis1987AFoundation}
\bibfield{author}{\bibinfo{person}{Gerardine DeSanctis} {and}
  \bibinfo{person}{R.~Brent Gallupe}.} \bibinfo{year}{1987}\natexlab{}.
\newblock \showarticletitle{A Foundation for the Study of Group Decision
  Support Systems}.
\newblock \bibinfo{journal}{\emph{Manage. Sci.}} \bibinfo{volume}{33},
  \bibinfo{number}{5} (\bibinfo{year}{1987}), \bibinfo{pages}{589–609}.
\newblock
\showISSN{0025-1909}


\bibitem[Eden(2020)]%
        {Eden2020}
\bibfield{author}{\bibinfo{person}{Colin Eden}.}
  \bibinfo{year}{2020}\natexlab{}.
\newblock \bibinfo{booktitle}{\emph{Behavioral Considerations in Group
  Support}}.
\newblock \bibinfo{publisher}{Springer International Publishing},
  \bibinfo{address}{Cham}, \bibinfo{pages}{1--16}.
\newblock
\urldef\tempurl%
\url{https://doi.org/10.1007/978-3-030-12051-1_34-1}
\showDOI{\tempurl}


\bibitem[Felfernig et~al\mbox{.}(2018)]%
        {Felfernig2018GRSIntro}
\bibfield{author}{\bibinfo{person}{Alexander Felfernig},
  \bibinfo{person}{Ludovico Boratto}, \bibinfo{person}{Martin Stettinger},
  {and} \bibinfo{person}{Marko Tkali}.} \bibinfo{year}{2018}\natexlab{}.
\newblock \bibinfo{booktitle}{\emph{Group Recommender Systems: An Introduction}
  (\bibinfo{edition}{1st} ed.)}.
\newblock \bibinfo{publisher}{Springer}.
\newblock


\bibitem[Felfernig et~al\mbox{.}(2012)]%
        {Felfernig2012Requirements}
\bibfield{author}{\bibinfo{person}{Alexander Felfernig},
  \bibinfo{person}{Christoph Zehentner}, \bibinfo{person}{Gerald Ninaus},
  \bibinfo{person}{Harald Grabner}, \bibinfo{person}{Walid Maalej},
  \bibinfo{person}{Dennis Pagano}, \bibinfo{person}{Leopold Weninger}, {and}
  \bibinfo{person}{Florian Reinfrank}.} \bibinfo{year}{2012}\natexlab{}.
\newblock \showarticletitle{Group Decision Support for Requirements
  Negotiation}. In \bibinfo{booktitle}{\emph{Advances in User Modeling}}.
  \bibinfo{pages}{105--116}.
\newblock


\bibitem[Feng et~al\mbox{.}(2023)]%
        {feng2023largelanguagemodelenhanced}
\bibfield{author}{\bibinfo{person}{Yue Feng}, \bibinfo{person}{Shuchang Liu},
  \bibinfo{person}{Zhenghai Xue}, \bibinfo{person}{Qingpeng Cai},
  \bibinfo{person}{Lantao Hu}, \bibinfo{person}{Peng Jiang},
  \bibinfo{person}{Kun Gai}, {and} \bibinfo{person}{Fei Sun}.}
  \bibinfo{year}{2023}\natexlab{}.
\newblock \bibinfo{title}{A Large Language Model Enhanced Conversational
  Recommender System}.
\newblock
\newblock
\showeprint[arxiv]{2308.06212}~[cs.IR]
\urldef\tempurl%
\url{https://arxiv.org/abs/2308.06212}
\showURL{%
\tempurl}


\bibitem[Fischer(2023)]%
        {fischer2023reflectivelinguisticprogrammingrlp}
\bibfield{author}{\bibinfo{person}{Kevin~A. Fischer}.}
  \bibinfo{year}{2023}\natexlab{}.
\newblock \bibinfo{title}{Reflective Linguistic Programming (RLP): A Stepping
  Stone in Socially-Aware AGI (SocialAGI)}.
\newblock
\newblock
\showeprint[arxiv]{2305.12647}~[cs.AI]
\urldef\tempurl%
\url{https://arxiv.org/abs/2305.12647}
\showURL{%
\tempurl}


\bibitem[Forsyth(2019)]%
        {Forsyth2019}
\bibfield{author}{\bibinfo{person}{D.R. Forsyth}.}
  \bibinfo{year}{2019}\natexlab{}.
\newblock \bibinfo{booktitle}{\emph{Group Dynamics} (\bibinfo{edition}{7}
  ed.)}.
\newblock \bibinfo{publisher}{Cengage}.
\newblock


\bibitem[Foster and Stefik(1986)]%
        {Foster1986Cognoter}
\bibfield{author}{\bibinfo{person}{Gregg Foster} {and} \bibinfo{person}{Mark
  Stefik}.} \bibinfo{year}{1986}\natexlab{}.
\newblock \showarticletitle{Cognoter: theory and practice of a colab-orative
  tool}. In \bibinfo{booktitle}{\emph{Proceedings of the 1986 ACM conference on
  Computer-supported cooperative work}}. \bibinfo{pages}{7--15}.
\newblock


\bibitem[Friedman et~al\mbox{.}(2023)]%
        {friedman2023leveraginglargelanguagemodels}
\bibfield{author}{\bibinfo{person}{Luke Friedman}, \bibinfo{person}{Sameer
  Ahuja}, \bibinfo{person}{David Allen}, \bibinfo{person}{Zhenning Tan},
  \bibinfo{person}{Hakim Sidahmed}, \bibinfo{person}{Changbo Long},
  \bibinfo{person}{Jun Xie}, \bibinfo{person}{Gabriel Schubiner},
  \bibinfo{person}{Ajay Patel}, \bibinfo{person}{Harsh Lara},
  \bibinfo{person}{Brian Chu}, \bibinfo{person}{Zexi Chen}, {and}
  \bibinfo{person}{Manoj Tiwari}.} \bibinfo{year}{2023}\natexlab{}.
\newblock \bibinfo{title}{Leveraging Large Language Models in Conversational
  Recommender Systems}.
\newblock
\newblock
\showeprint[arxiv]{2305.07961}~[cs.IR]
\urldef\tempurl%
\url{https://arxiv.org/abs/2305.07961}
\showURL{%
\tempurl}


\bibitem[Govers et~al\mbox{.}(2024)]%
        {Govers2024Mediation}
\bibfield{author}{\bibinfo{person}{Jarod Govers}, \bibinfo{person}{Eduardo
  Velloso}, \bibinfo{person}{Vassilis Kostakos}, {and} \bibinfo{person}{Jorge
  Goncalves}.} \bibinfo{year}{2024}\natexlab{}.
\newblock \showarticletitle{AI-driven mediation strategies for audience
  depolarisation in online debates}. In \bibinfo{booktitle}{\emph{Proceedings
  of the 2024 CHI Conference on Human Factors in Computing Systems}}.
  \bibinfo{pages}{1--18}.
\newblock


\bibitem[Guan et~al\mbox{.}(2025)]%
        {guan2025evaluatingllmbasedagentsmultiturn}
\bibfield{author}{\bibinfo{person}{Shengyue Guan}, \bibinfo{person}{Haoyi
  Xiong}, \bibinfo{person}{Jindong Wang}, \bibinfo{person}{Jiang Bian},
  \bibinfo{person}{Bin Zhu}, {and} \bibinfo{person}{Jian guang Lou}.}
  \bibinfo{year}{2025}\natexlab{}.
\newblock \bibinfo{title}{Evaluating LLM-based Agents for Multi-Turn
  Conversations: A Survey}.
\newblock
\newblock
\showeprint[arxiv]{2503.22458}~[cs.CL]
\urldef\tempurl%
\url{https://arxiv.org/abs/2503.22458}
\showURL{%
\tempurl}


\bibitem[Gurkan and Yan({[n.\,d.]})]%
        {Gurkan2023ChatbotCata}
\bibfield{author}{\bibinfo{person}{Necdet Gurkan} {and} \bibinfo{person}{Bei
  Yan}.} \bibinfo{year}{[n.\,d.]}\natexlab{}.
\newblock \showarticletitle{Chatbot Catalysts: Improving Team Decision-Making
  Through Cognitive Diversity and Information Elaboration}. In
  \bibinfo{booktitle}{\emph{ICIS 2023 Proceedings}}.
\newblock
\urldef\tempurl%
\url{https://par.nsf.gov/biblio/10545067}
\showURL{%
\tempurl}


\bibitem[Guzzi et~al\mbox{.}(2011)]%
        {Guzzi2011Interactive}
\bibfield{author}{\bibinfo{person}{Francesca Guzzi}, \bibinfo{person}{Francesco
  Ricci}, {and} \bibinfo{person}{Robin Burke}.}
  \bibinfo{year}{2011}\natexlab{}.
\newblock \showarticletitle{Interactive multi-party critiquing for group
  recommendation}. In \bibinfo{booktitle}{\emph{Proceedings of the Fifth ACM
  Conference on Recommender Systems}} \emph{(\bibinfo{series}{RecSys '11})}.
  \bibinfo{pages}{265–268}.
\newblock
\urldef\tempurl%
\url{https://doi.org/10.1145/2043932.2043980}
\showDOI{\tempurl}


\bibitem[He et~al\mbox{.}(2023)]%
        {he2023large}
\bibfield{author}{\bibinfo{person}{Zhankui He}, \bibinfo{person}{Zhouhang Xie},
  \bibinfo{person}{Rahul Jha}, \bibinfo{person}{Harald Steck},
  \bibinfo{person}{Dawen Liang}, \bibinfo{person}{Yesu Feng},
  \bibinfo{person}{Bodhisattwa~Prasad Majumder}, \bibinfo{person}{Nathan
  Kallus}, {and} \bibinfo{person}{Julian Mcauley}.}
  \bibinfo{year}{2023}\natexlab{}.
\newblock \showarticletitle{Large Language Models as Zero-Shot Conversational
  Recommenders}. In \bibinfo{booktitle}{\emph{Proceedings of the 32nd ACM
  International Conference on Information and Knowledge Management}}
  \emph{(\bibinfo{series}{CIKM '23})}. \bibinfo{pages}{720–730}.
\newblock
\urldef\tempurl%
\url{https://doi.org/10.1145/3583780.3614949}
\showDOI{\tempurl}


\bibitem[Hitzler and Sarker(2022)]%
        {hitzler2022neurosymbolic}
\bibfield{editor}{\bibinfo{person}{Pascal Hitzler} {and}
  \bibinfo{person}{Md~Kamruzzaman Sarker}} (Eds.).
  \bibinfo{year}{2022}\natexlab{}.
\newblock \bibinfo{booktitle}{\emph{Neuro-Symbolic Artificial Intelligence: The
  State of the Art}}. \bibinfo{series}{Frontiers in Artificial Intelligence and
  Applications}, Vol.~\bibinfo{volume}{342}.
\newblock \bibinfo{publisher}{IOS Press}, \bibinfo{address}{Amsterdam}.
\newblock
\showISBNx{978-1-64368-244-0}
\urldef\tempurl%
\url{https://doi.org/10.3233/FAIA342}
\showDOI{\tempurl}


\bibitem[Hou et~al\mbox{.}({[n.\,d.]})]%
        {hou2024large}
\bibfield{author}{\bibinfo{person}{Yupeng Hou}, \bibinfo{person}{Junjie Zhang},
  \bibinfo{person}{Zihan Lin}, \bibinfo{person}{Hongyu Lu},
  \bibinfo{person}{Ruobing Xie}, \bibinfo{person}{Julian McAuley}, {and}
  \bibinfo{person}{Wayne~Xin Zhao}.} \bibinfo{year}{[n.\,d.]}\natexlab{}.
\newblock \showarticletitle{Large Language Models are Zero-Shot Rankers for
  ecommender Systems}. In \bibinfo{booktitle}{\emph{Advances in Information
  Retrieval: 46th European Conference on Information Retrieval, ECIR 2024}}.
  \bibinfo{pages}{364–381}.
\newblock
\urldef\tempurl%
\url{https://doi.org/10.1007/978-3-031-56060-6_24}
\showDOI{\tempurl}


\bibitem[Hou et~al\mbox{.}(2023)]%
        {hou2023large}
\bibfield{author}{\bibinfo{person}{Yupeng Hou}, \bibinfo{person}{Junjie Zhang},
  \bibinfo{person}{Zihan Lin}, \bibinfo{person}{Hongyu Lu},
  \bibinfo{person}{Ruobing Xie}, \bibinfo{person}{Julian McAuley}, {and}
  \bibinfo{person}{Wayne~Xin Zhao}.} \bibinfo{year}{2023}\natexlab{}.
\newblock \bibinfo{title}{Large Language Models are Zero-Shot Rankers for
  Recommender Systems}.
\newblock
\newblock
\showeprint[arxiv]{2305.08845}~[cs.IR]


\bibitem[Huang et~al\mbox{.}(2024)]%
        {huang2024understandingplanningllmagents}
\bibfield{author}{\bibinfo{person}{Xu Huang}, \bibinfo{person}{Weiwen Liu},
  \bibinfo{person}{Xiaolong Chen}, \bibinfo{person}{Xingmei Wang},
  \bibinfo{person}{Hao Wang}, \bibinfo{person}{Defu Lian},
  \bibinfo{person}{Yasheng Wang}, \bibinfo{person}{Ruiming Tang}, {and}
  \bibinfo{person}{Enhong Chen}.} \bibinfo{year}{2024}\natexlab{}.
\newblock \bibinfo{title}{Understanding the planning of LLM agents: A survey}.
\newblock
\newblock
\showeprint[arxiv]{2402.02716}~[cs.AI]
\urldef\tempurl%
\url{https://arxiv.org/abs/2402.02716}
\showURL{%
\tempurl}


\bibitem[Iovine et~al\mbox{.}(2020)]%
        {Iovine2020Conversational}
\bibfield{author}{\bibinfo{person}{Andrea Iovine}, \bibinfo{person}{Fedelucio
  Narducci}, {and} \bibinfo{person}{Giovanni Semeraro}.}
  \bibinfo{year}{2020}\natexlab{}.
\newblock \showarticletitle{Conversational Recommender Systems and natural
  language: : {A} study through the ConveRSE framework}.
\newblock \bibinfo{journal}{\emph{Decis. Support Syst.}}  \bibinfo{volume}{131}
  (\bibinfo{year}{2020}), \bibinfo{pages}{113250}.
\newblock
\urldef\tempurl%
\url{https://doi.org/10.1016/J.DSS.2020.113250}
\showDOI{\tempurl}


\bibitem[Jameson and Smyth(2007)]%
        {Jameson2007}
\bibfield{author}{\bibinfo{person}{Anthony Jameson} {and}
  \bibinfo{person}{Barry Smyth}.} \bibinfo{year}{2007}\natexlab{}.
\newblock \bibinfo{booktitle}{\emph{Recommendation to Groups}}.
\newblock \bibinfo{publisher}{Springer Berlin Heidelberg},
  \bibinfo{pages}{596--627}.
\newblock
\urldef\tempurl%
\url{https://doi.org/10.1007/978-3-540-72079-9_20}
\showDOI{\tempurl}


\bibitem[Jameson et~al\mbox{.}(2022)]%
        {Jameson2022}
\bibfield{author}{\bibinfo{person}{Anthony Jameson},
  \bibinfo{person}{Martijn~C. Willemsen}, {and} \bibinfo{person}{Alexander
  Felfernig}.} \bibinfo{year}{2022}\natexlab{}.
\newblock \bibinfo{booktitle}{\emph{Individual and Group Decision Making and
  Recommender Systems}}.
\newblock \bibinfo{publisher}{Springer US}, \bibinfo{address}{New York, NY},
  \bibinfo{pages}{789--832}.
\newblock
\showISBNx{978-1-0716-2197-4}
\urldef\tempurl%
\url{https://doi.org/10.1007/978-1-0716-2197-4_21}
\showDOI{\tempurl}


\bibitem[Jannach(2022)]%
        {jannach2022evaluatingcrs}
\bibfield{author}{\bibinfo{person}{Dietmar Jannach}.}
  \bibinfo{year}{2022}\natexlab{}.
\newblock \showarticletitle{Evaluating Conversational Recommender Systems}.
\newblock \bibinfo{journal}{\emph{Artificial Intelligence Review}}
  \bibinfo{volume}{56} (\bibinfo{year}{2022}), \bibinfo{pages}{2365–2400}.
\newblock


\bibitem[Jannach and Chen(2022)]%
        {jannach2022grandchallenge}
\bibfield{author}{\bibinfo{person}{Dietmar Jannach} {and} \bibinfo{person}{Li
  Chen}.} \bibinfo{year}{2022}\natexlab{}.
\newblock \showarticletitle{Conversational recommendation: A grand AI
  challenge}.
\newblock \bibinfo{journal}{\emph{AI Magazine}} \bibinfo{volume}{43},
  \bibinfo{number}{2} (\bibinfo{year}{2022}), \bibinfo{pages}{151--163}.
\newblock
\urldef\tempurl%
\url{https://doi.org/10.1002/aaai.12059}
\showDOI{\tempurl}


\bibitem[Jannach et~al\mbox{.}(2021)]%
        {jannach2021crscsur}
\bibfield{author}{\bibinfo{person}{Dietmar Jannach}, \bibinfo{person}{Ahtsham
  Manzoor}, \bibinfo{person}{Wanling Cai}, {and} \bibinfo{person}{Li Chen}.}
  \bibinfo{year}{2021}\natexlab{}.
\newblock \showarticletitle{A Survey on Conversational Recommender Systems}.
\newblock \bibinfo{journal}{\emph{Comput. Surveys}} \bibinfo{volume}{54},
  \bibinfo{number}{5} (\bibinfo{year}{2021}), \bibinfo{pages}{1--26}.
\newblock


\bibitem[Jin et~al\mbox{.}(2024)]%
        {Jin2024crsque}
\bibfield{author}{\bibinfo{person}{Yucheng Jin}, \bibinfo{person}{Li Chen},
  \bibinfo{person}{Wanling Cai}, {and} \bibinfo{person}{Xianglin Zhao}.}
  \bibinfo{year}{2024}\natexlab{}.
\newblock \showarticletitle{CRS-Que: A User-centric Evaluation Framework for
  Conversational Recommender Systems}.
\newblock \bibinfo{journal}{\emph{ACM Trans. Recomm. Syst.}}
  \bibinfo{volume}{2}, \bibinfo{number}{1}, Article \bibinfo{articleno}{2}
  (\bibinfo{date}{March} \bibinfo{year}{2024}).
\newblock
\urldef\tempurl%
\url{https://doi.org/10.1145/3631534}
\showDOI{\tempurl}


\bibitem[Jr. et~al\mbox{.}(1991)]%
        {Nunamaker1991Electronig}
\bibfield{author}{\bibinfo{person}{Jay F.~Nunamaker Jr.},
  \bibinfo{person}{Alan~R. Dennis}, \bibinfo{person}{Joseph~S. Valacich},
  \bibinfo{person}{Douglas~R. Vogel}, {and} \bibinfo{person}{Joey~F. George}.}
  \bibinfo{year}{1991}\natexlab{}.
\newblock \showarticletitle{Electronic Meeting Systems To Support Group Work}.
\newblock \bibinfo{journal}{\emph{Commun. {ACM}}} \bibinfo{volume}{34},
  \bibinfo{number}{7} (\bibinfo{year}{1991}), \bibinfo{pages}{40--61}.
\newblock
\urldef\tempurl%
\url{https://doi.org/10.1145/105783.105793}
\showDOI{\tempurl}


\bibitem[Karahodža et~al\mbox{.}(2025)]%
        {Karahodza2025GroupDynamics}
\bibfield{author}{\bibinfo{person}{Esma Karahodža}, \bibinfo{person}{Amra
  Delić}, {and} \bibinfo{person}{Francesco Ricci}.}
  \bibinfo{year}{2025}\natexlab{}.
\newblock \showarticletitle{Conceptual Framework for Group Dynamics Modeling
  from Group Chat Interactions}. In \bibinfo{booktitle}{\emph{Adjunct
  Proceedings of the 33rd ACM Conference on User Modeling, Adaptation and
  Personalization (UMAP Adjunct '25)}}. \bibinfo{pages}{5}.
\newblock
\urldef\tempurl%
\url{https://doi.org/10.1145/3708319.3733682}
\showDOI{\tempurl}


\bibitem[Keestra et~al\mbox{.}(2016)]%
        {Keestra_Menken_2016}
\bibfield{author}{\bibinfo{person}{Machiel Keestra}, \bibinfo{person}{Lucas
  Rutting}, \bibinfo{person}{Ger Post}, \bibinfo{person}{Mieke de Roo},
  \bibinfo{person}{Sylvia Blad}, {and} \bibinfo{person}{Linda de Greef}.}
  \bibinfo{year}{2016}\natexlab{}.
\newblock \bibinfo{booktitle}{\emph{An Introduction to Interdisciplinary
  Research: Theory and Practice}}.
\newblock \bibinfo{publisher}{Amsterdam University Press}.
\newblock


\bibitem[Kilgour and Eden(2019)]%
        {kilgour2019handbook}
\bibfield{editor}{\bibinfo{person}{D.~Marc Kilgour} {and}
  \bibinfo{person}{Colin Eden}} (Eds.). \bibinfo{year}{2019}\natexlab{}.
\newblock \bibinfo{booktitle}{\emph{Handbook of Group Decision and
  Negotiation}}.
\newblock \bibinfo{publisher}{Springer}.
\newblock
\showISBNx{978-3-030-12050-4}
\urldef\tempurl%
\url{https://doi.org/10.1007/978-3-030-12051-1}
\showDOI{\tempurl}


\bibitem[Kim et~al\mbox{.}(2020)]%
        {Kim2020Bot}
\bibfield{author}{\bibinfo{person}{Soomin Kim}, \bibinfo{person}{Jinsu Eun},
  \bibinfo{person}{Changhoon Oh}, \bibinfo{person}{Bongwon Suh}, {and}
  \bibinfo{person}{Joonhwan Lee}.} \bibinfo{year}{2020}\natexlab{}.
\newblock \showarticletitle{Bot in the Bunch: Facilitating Group Chat
  Discussion by Improving Efficiency and Participation with a Chatbot}. In
  \bibinfo{booktitle}{\emph{Proceedings of the 2020 CHI Conference on Human
  Factors in Computing Systems}} (Honolulu, HI, USA)
  \emph{(\bibinfo{series}{CHI '20})}. \bibinfo{publisher}{Association for
  Computing Machinery}, \bibinfo{address}{New York, NY, USA},
  \bibinfo{pages}{1–13}.
\newblock
\urldef\tempurl%
\url{https://doi.org/10.1145/3313831.3376785}
\showDOI{\tempurl}


\bibitem[Kim et~al\mbox{.}(2021)]%
        {Kim2021ModeratorChatbot}
\bibfield{author}{\bibinfo{person}{Soomin Kim}, \bibinfo{person}{Jinsu Eun},
  \bibinfo{person}{Joseph Seering}, {and} \bibinfo{person}{Joonhwan Lee}.}
  \bibinfo{year}{2021}\natexlab{}.
\newblock \showarticletitle{Moderator Chatbot for Deliberative Discussion:
  Effects of Discussion Structure and Discussant Facilitation}.
\newblock \bibinfo{journal}{\emph{Proc. {ACM} Hum. Comput. Interact.}}
  \bibinfo{volume}{5}, \bibinfo{number}{{CSCW1}} (\bibinfo{year}{2021}),
  \bibinfo{pages}{87:1--87:26}.
\newblock
\urldef\tempurl%
\url{https://doi.org/10.1145/3449161}
\showDOI{\tempurl}


\bibitem[Kim et~al\mbox{.}(2025)]%
        {kim-etal-2025-stop}
\bibfield{author}{\bibinfo{person}{SungHwan Kim}, \bibinfo{person}{Kwangwook
  Seo}, \bibinfo{person}{Tongyoung Kim}, \bibinfo{person}{Jinyoung Yeo}, {and}
  \bibinfo{person}{Dongha Lee}.} \bibinfo{year}{2025}\natexlab{}.
\newblock \showarticletitle{Stop Playing the Guessing Game! Evaluating
  Conversational Recommender Systems via Target-free User Simulation}. In
  \bibinfo{booktitle}{\emph{Findings of the Association for Computational
  Linguistics: EMNLP 2025}}. \bibinfo{address}{Suzhou, China},
  \bibinfo{pages}{19588--19605}.
\newblock
\urldef\tempurl%
\url{https://doi.org/10.18653/v1/2025.findings-emnlp.1067}
\showDOI{\tempurl}


\bibitem[Kuhail et~al\mbox{.}(2024)]%
        {kuhal2024polyadic}
\bibfield{author}{\bibinfo{person}{Mohammad~Amin Kuhail},
  \bibinfo{person}{Imran Taj}, \bibinfo{person}{Saifeddin Alimamy}, {and}
  \bibinfo{person}{Bayan~Abu Shawar}.} \bibinfo{year}{2024}\natexlab{}.
\newblock \showarticletitle{A review on polyadic chatbots: trends, challenges,
  and future research directions}.
\newblock \bibinfo{journal}{\emph{Knowledge and Information Systems}}
  \bibinfo{volume}{online first} (\bibinfo{year}{2024}).
\newblock
\urldef\tempurl%
\url{https://doi.org/10.1007/s10115-024-02287-0}
\showDOI{\tempurl}


\bibitem[Lee et~al\mbox{.}(2025a)]%
        {Lee2025Catalysts}
\bibfield{author}{\bibinfo{person}{Soohwan Lee}, \bibinfo{person}{Seoyeong
  Hwang}, \bibinfo{person}{Dajung Kim}, {and} \bibinfo{person}{Kyungho Lee}.}
  \bibinfo{year}{2025}\natexlab{a}.
\newblock \showarticletitle{Conversational agents as catalysts for critical
  thinking: Challenging social influence in group decision-making}. In
  \bibinfo{booktitle}{\emph{Proceedings of the Extended Abstracts of the CHI
  Conference on Human Factors in Computing Systems}}. \bibinfo{pages}{1--12}.
\newblock


\bibitem[Lee et~al\mbox{.}(2025b)]%
        {Lee2025Amplifying}
\bibfield{author}{\bibinfo{person}{SooHwan Lee}, \bibinfo{person}{Mingyu Kim},
  \bibinfo{person}{Seoyeong Hwang}, \bibinfo{person}{Dajung Kim}, {and}
  \bibinfo{person}{Kyungho Lee}.} \bibinfo{year}{2025}\natexlab{b}.
\newblock \showarticletitle{Amplifying minority voices: AI-mediated devil's
  advocate system for inclusive group decision-making}. In
  \bibinfo{booktitle}{\emph{Companion Proceedings of the 30th International
  Conference on Intelligent User Interfaces}}. \bibinfo{pages}{17--21}.
\newblock


\bibitem[Lee et~al\mbox{.}(2020)]%
        {Lee2020SolutionChat}
\bibfield{author}{\bibinfo{person}{Sung-Chul Lee}, \bibinfo{person}{Jaeyoon
  Song}, \bibinfo{person}{Eun-Young Ko}, \bibinfo{person}{Seongho Park},
  \bibinfo{person}{Jihee Kim}, {and} \bibinfo{person}{Juho Kim}.}
  \bibinfo{year}{2020}\natexlab{}.
\newblock \showarticletitle{SolutionChat: Real-time Moderator Support for
  Chat-based Structured Discussion}. In \bibinfo{booktitle}{\emph{CHI '20}}.
  \bibinfo{pages}{1--12}.
\newblock
\urldef\tempurl%
\url{https://doi.org/10.1145/3313831.3376609}
\showDOI{\tempurl}


\bibitem[Leviathan and Matias(2018)]%
        {GoogleDuplex2018}
\bibfield{author}{\bibinfo{person}{Yaniv Leviathan} {and}
  \bibinfo{person}{Yossi Matias}.} \bibinfo{year}{2018}\natexlab{}.
\newblock \bibinfo{title}{Google Duplex: An AI System for Accomplishing
  Real-World Tasks Over the Phone}.
\newblock
\newblock
\urldef\tempurl%
\url{https://ai.googleblog.com/2018/05/duplex-ai-system-for-natural-conversation.html}
\showURL{%
\tempurl}


\bibitem[Lex et~al\mbox{.}(2021)]%
        {Lex2021Psy}
\bibfield{author}{\bibinfo{person}{Elisabeth Lex}, \bibinfo{person}{Dominik
  Kowald}, \bibinfo{person}{Paul Seitlinger}, \bibinfo{person}{Thi Ngoc~Trang
  Tran}, \bibinfo{person}{Alexander Felfernig}, {and} \bibinfo{person}{Markus
  Schedl}.} \bibinfo{year}{2021}\natexlab{}.
\newblock \showarticletitle{Psychology-informed Recommender Systems}.
\newblock \bibinfo{journal}{\emph{Foundations and Trends® in Information
  Retrieval}} \bibinfo{volume}{15}, \bibinfo{number}{2} (\bibinfo{year}{2021}),
  \bibinfo{pages}{134--242}.
\newblock
\urldef\tempurl%
\url{https://doi.org/10.1561/1500000090}
\showDOI{\tempurl}


\bibitem[Li(2018)]%
        {Xitong2018Impact}
\bibfield{author}{\bibinfo{person}{Xitong Li}.}
  \bibinfo{year}{2018}\natexlab{}.
\newblock \showarticletitle{Impact of Average Rating on Social Media
  Endorsement: The Moderating Role of Rating Dispersion and Discount
  Threshold}.
\newblock \bibinfo{journal}{\emph{Info. Sys. Research}} \bibinfo{volume}{29},
  \bibinfo{number}{3} (\bibinfo{year}{2018}), \bibinfo{pages}{739–754}.
\newblock
\urldef\tempurl%
\url{https://doi.org/10.1287/isre.2017.0728}
\showDOI{\tempurl}


\bibitem[Liu et~al\mbox{.}(2023)]%
        {liu2023chatgpt}
\bibfield{author}{\bibinfo{person}{Junling Liu}, \bibinfo{person}{Chao Liu},
  \bibinfo{person}{Renjie Lv}, \bibinfo{person}{Kang Zhou}, {and}
  \bibinfo{person}{Yan Zhang}.} \bibinfo{year}{2023}\natexlab{}.
\newblock \bibinfo{title}{Is ChatGPT a Good Recommender? A Preliminary Study}.
\newblock
\newblock
\showeprint[arxiv]{2304.10149}~[cs.IR]


\bibitem[Mahmood and Ricci(2009)]%
        {Tariq2009Improving}
\bibfield{author}{\bibinfo{person}{Tariq Mahmood} {and}
  \bibinfo{person}{Francesco Ricci}.} \bibinfo{year}{2009}\natexlab{}.
\newblock \showarticletitle{Improving recommender systems with adaptive
  conversational strategies}. In \bibinfo{booktitle}{\emph{Proceedings of the
  20th ACM Conference on Hypertext and Hypermedia}} \emph{(\bibinfo{series}{HT
  '09})}. \bibinfo{pages}{73–82}.
\newblock
\urldef\tempurl%
\url{https://doi.org/10.1145/1557914.1557930}
\showDOI{\tempurl}


\bibitem[Manzoor et~al\mbox{.}(2024)]%
        {Manzoor2024Chatgpt}
\bibfield{author}{\bibinfo{person}{Ahtsham Manzoor}, \bibinfo{person}{Samuel~C.
  Ziegler}, \bibinfo{person}{Klaus Maria.~Pirker Garcia}, {and}
  \bibinfo{person}{Dietmar Jannach}.} \bibinfo{year}{2024}\natexlab{}.
\newblock \showarticletitle{ChatGPT as a Conversational Recommender System: A
  User-Centric Analysis}. In \bibinfo{booktitle}{\emph{Proceedings of the 32nd
  ACM Conference on User Modeling, Adaptation and Personalization}} (Cagliari,
  Italy) \emph{(\bibinfo{series}{UMAP '24})}. \bibinfo{publisher}{Association
  for Computing Machinery}, \bibinfo{address}{New York, NY, USA},
  \bibinfo{pages}{267–272}.
\newblock
\urldef\tempurl%
\url{https://doi.org/10.1145/3627043.3659574}
\showDOI{\tempurl}


\bibitem[Martinovski(2021)]%
        {Martinovski2021}
\bibfield{author}{\bibinfo{person}{Bilyana Martinovski}.}
  \bibinfo{year}{2021}\natexlab{}.
\newblock \showarticletitle{Role of Emotion in Group Decision and Negotiation}.
\newblock In \bibinfo{booktitle}{\emph{Handbook of Group Decision and
  Negotiation}}, \bibfield{editor}{\bibinfo{person}{D.~Marc Kilgour} {and}
  \bibinfo{person}{Colin Eden}} (Eds.). \bibinfo{publisher}{Springer
  International Publishing}, \bibinfo{pages}{157--192}.
\newblock
\urldef\tempurl%
\url{https://doi.org/10.1007/978-3-030-49629-6\_5}
\showDOI{\tempurl}


\bibitem[Masthoff(2015)]%
        {Masthoff2015GRS}
\bibfield{author}{\bibinfo{person}{Judith Masthoff}.}
  \bibinfo{year}{2015}\natexlab{}.
\newblock \bibinfo{booktitle}{\emph{Group Recommender Systems: Aggregation,
  Satisfaction and Group Attributes}}.
\newblock \bibinfo{publisher}{Springer US}, \bibinfo{pages}{743--776}.
\newblock
\urldef\tempurl%
\url{https://doi.org/10.1007/978-1-4899-7637-6_22}
\showDOI{\tempurl}


\bibitem[Masthoff and Delic(2022)]%
        {Masthoff2022GRSBeyond}
\bibfield{author}{\bibinfo{person}{Judith Masthoff} {and} \bibinfo{person}{Amra
  Delic}.} \bibinfo{year}{2022}\natexlab{}.
\newblock \showarticletitle{Group Recommender Systems: Beyond Preference
  Aggregation}.
\newblock In \bibinfo{booktitle}{\emph{Recommender Systems Handbook}},
  \bibfield{editor}{\bibinfo{person}{Francesco Ricci}, \bibinfo{person}{Lior
  Rokach}, {and} \bibinfo{person}{Bracha Shapira}} (Eds.).
  \bibinfo{publisher}{Springer {US}}, \bibinfo{pages}{381--420}.
\newblock
\urldef\tempurl%
\url{https://doi.org/10.1007/978-1-0716-2197-4\_10}
\showDOI{\tempurl}


\bibitem[Masthoff and Gatt(2006)]%
        {MasthoffG06}
\bibfield{author}{\bibinfo{person}{Judith Masthoff} {and}
  \bibinfo{person}{Albert Gatt}.} \bibinfo{year}{2006}\natexlab{}.
\newblock \showarticletitle{In pursuit of satisfaction and the prevention of
  embarrassment: affective state in group recommender systems}.
\newblock \bibinfo{journal}{\emph{User Model. User Adapt. Interact.}}
  \bibinfo{volume}{16}, \bibinfo{number}{3-4} (\bibinfo{year}{2006}),
  \bibinfo{pages}{281--319}.
\newblock
\urldef\tempurl%
\url{https://doi.org/10.1007/S11257-006-9008-3}
\showDOI{\tempurl}


\bibitem[McCarthy and Anagnost(1998)]%
        {McCarthy1998MusicFX}
\bibfield{author}{\bibinfo{person}{Joseph~F. McCarthy} {and}
  \bibinfo{person}{Theodore~D. Anagnost}.} \bibinfo{year}{1998}\natexlab{}.
\newblock \showarticletitle{MusicFX: an arbiter of group preferences for
  computer supported collaborative workouts}. In
  \bibinfo{booktitle}{\emph{Proceedings of the 1998 ACM Conference on Computer
  Supported Cooperative Work}} \emph{(\bibinfo{series}{CSCW '98})}.
  \bibinfo{pages}{363–372}.
\newblock
\urldef\tempurl%
\url{https://doi.org/10.1145/289444.289511}
\showDOI{\tempurl}


\bibitem[McCarthy et~al\mbox{.}(2006)]%
        {McCarthy2006Needs}
\bibfield{author}{\bibinfo{person}{Kevin McCarthy}, \bibinfo{person}{Lorraine
  McGinty}, \bibinfo{person}{Barry Smyth}, {and} \bibinfo{person}{Maria
  Salam{\'o}}.} \bibinfo{year}{2006}\natexlab{}.
\newblock \showarticletitle{The Needs of the Many: A Case-Based Group
  Recommender System}. In \bibinfo{booktitle}{\emph{Advances in Case-Based
  Reasoning}}. \bibinfo{pages}{196--210}.
\newblock


\bibitem[Najafian et~al\mbox{.}(2021)]%
        {Najafian2021Factors}
\bibfield{author}{\bibinfo{person}{Shabnam Najafian}, \bibinfo{person}{Amra
  Delic}, \bibinfo{person}{Marko Tkalcic}, {and} \bibinfo{person}{Nava
  Tintarev}.} \bibinfo{year}{2021}\natexlab{}.
\newblock \showarticletitle{Factors Influencing Privacy Concern for
  Explanations of Group Recommendation}. In
  \bibinfo{booktitle}{\emph{Proceedings of the 29th ACM Conference on User
  Modeling, Adaptation and Personalization}} \emph{(\bibinfo{series}{UMAP
  '21})}. \bibinfo{pages}{14–23}.
\newblock
\urldef\tempurl%
\url{https://doi.org/10.1145/3450613.3456845}
\showDOI{\tempurl}


\bibitem[Nguyen and Ricci(2017)]%
        {Nguyen2017AChatBased}
\bibfield{author}{\bibinfo{person}{Ngoc Nguyen} {and}
  \bibinfo{person}{Francesco Ricci}.} \bibinfo{year}{2017}\natexlab{}.
\newblock \bibinfo{booktitle}{\emph{A Chat-Based Group Recommender System for
  Tourism}}.
\newblock \bibinfo{pages}{17--30}.
\newblock
\showISBNx{978-3-319-51167-2}
\urldef\tempurl%
\url{https://doi.org/10.1007/978-3-319-51168-9_2}
\showDOI{\tempurl}


\bibitem[Ni et~al\mbox{.}(2026)]%
        {ni-etal-2026-survey}
\bibfield{author}{\bibinfo{person}{Bo Ni}, \bibinfo{person}{Yu Wang},
  \bibinfo{person}{Leyao Wang}, \bibinfo{person}{Branislav Kveton},
  \bibinfo{person}{Franck Dernoncourt}, \bibinfo{person}{Yu Xia},
  \bibinfo{person}{Hongjie Chen}, \bibinfo{person}{Reuben Luera},
  \bibinfo{person}{Samyadeep Basu}, \bibinfo{person}{Subhojyoti Mukherjee},
  \bibinfo{person}{Puneet Mathur}, \bibinfo{person}{Nesreen~K. Ahmed},
  \bibinfo{person}{Junda Wu}, \bibinfo{person}{Li Li}, \bibinfo{person}{Huixin
  Zhang}, \bibinfo{person}{Ruiyi Zhang}, \bibinfo{person}{Tong Yu},
  \bibinfo{person}{Sungchul Kim}, \bibinfo{person}{Jiuxiang Gu},
  \bibinfo{person}{Zhengzhong Tu}, \bibinfo{person}{Alexa Siu},
  \bibinfo{person}{Zichao Wang}, \bibinfo{person}{Seunghyun Yoon},
  \bibinfo{person}{Nedim Lipka}, \bibinfo{person}{Namyong Park},
  \bibinfo{person}{Zihao Lin}, \bibinfo{person}{Trung Bui},
  \bibinfo{person}{Yue Zhao}, \bibinfo{person}{Tyler Derr}, {and}
  \bibinfo{person}{Ryan~A. Rossi}.} \bibinfo{year}{2026}\natexlab{}.
\newblock \showarticletitle{A Survey on {LLM}-based Conversational User
  Simulation}. In \bibinfo{booktitle}{\emph{Proceedings of the 19th Conference
  of the {E}uropean Chapter of the {A}ssociation for {C}omputational
  {L}inguistics (Volume 1: Long Papers)}}. \bibinfo{pages}{4266--4301}.
\newblock
\urldef\tempurl%
\url{https://doi.org/10.18653/v1/2026.eacl-long.200}
\showDOI{\tempurl}


\bibitem[Nunamaker et~al\mbox{.}(1996)]%
        {Nunamaker1996Lessons}
\bibfield{author}{\bibinfo{person}{Jay~F. Nunamaker},
  \bibinfo{person}{Robert~O. Briggs}, \bibinfo{person}{Daniel~D. Mittleman},
  \bibinfo{person}{Douglas~R. Vogel}, {and} \bibinfo{person}{Pierre~A.
  Balthazard}.} \bibinfo{year}{1996}\natexlab{}.
\newblock \showarticletitle{Lessons from a dozen years of group support systems
  research: a discussion of lab and field findings}.
\newblock \bibinfo{journal}{\emph{J. Manage. Inf. Syst.}} \bibinfo{volume}{13},
  \bibinfo{number}{3} (\bibinfo{year}{1996}), \bibinfo{pages}{163–207}.
\newblock
\showISSN{0742-1222}
\urldef\tempurl%
\url{https://doi.org/10.1080/07421222.1996.11518138}
\showDOI{\tempurl}


\bibitem[Nunes and Jannach(2017)]%
        {NunesJannachUmuai2017}
\bibfield{author}{\bibinfo{person}{Ingrid Nunes} {and} \bibinfo{person}{Dietmar
  Jannach}.} \bibinfo{year}{2017}\natexlab{}.
\newblock \showarticletitle{A Systematic Review and Taxonomy of Explanations in
  Decision Support and Recommender Systems}.
\newblock \bibinfo{journal}{\emph{User-Modeling and User-Adapted Interaction}}
  \bibinfo{volume}{27}, \bibinfo{number}{3--5} (\bibinfo{year}{2017}),
  \bibinfo{pages}{393--444}.
\newblock


\bibitem[O'Connor et~al\mbox{.}(2001)]%
        {OConnor2001PolyLens}
\bibfield{author}{\bibinfo{person}{Mark O'Connor}, \bibinfo{person}{Dan
  Cosley}, \bibinfo{person}{Joseph~A. Konstan}, {and} \bibinfo{person}{John
  Riedl}.} \bibinfo{year}{2001}\natexlab{}.
\newblock \showarticletitle{PolyLens: A Recommender System for Groups of
  Users}. In \bibinfo{booktitle}{\emph{ECSCW 2001: Proceedings of the Seventh
  European Conference on Computer Supported Cooperative Work}}.
  \bibinfo{pages}{199--218}.
\newblock
\urldef\tempurl%
\url{https://doi.org/10.1007/0-306-48019-0_11}
\showDOI{\tempurl}


\bibitem[Papachristou et~al\mbox{.}(2025)]%
        {Papachristou2025Leveraging}
\bibfield{author}{\bibinfo{person}{Marios Papachristou},
  \bibinfo{person}{Longqi Yang}, {and} \bibinfo{person}{Chin-Chia Hsu}.}
  \bibinfo{year}{2025}\natexlab{}.
\newblock \showarticletitle{Leveraging Large Language Models for Collective
  Decision-Making}.
\newblock \bibinfo{journal}{\emph{Proc. ACM Hum.-Comput. Interact.}}
  \bibinfo{volume}{9}, \bibinfo{number}{7}, Article
  \bibinfo{articleno}{CSCW237} (\bibinfo{year}{2025}).
\newblock
\urldef\tempurl%
\url{https://doi.org/10.1145/3757418}
\showDOI{\tempurl}


\bibitem[Peng et~al\mbox{.}(2025)]%
        {peng2025surveyllmpoweredagentsrecommender}
\bibfield{author}{\bibinfo{person}{Qiyao Peng}, \bibinfo{person}{Hongtao Liu},
  \bibinfo{person}{Hua Huang}, \bibinfo{person}{Qing Yang}, {and}
  \bibinfo{person}{Minglai Shao}.} \bibinfo{year}{2025}\natexlab{}.
\newblock \bibinfo{title}{A Survey on LLM-powered Agents for Recommender
  Systems}.
\newblock
\newblock
\showeprint[arxiv]{2502.10050}~[cs.IR]
\urldef\tempurl%
\url{https://arxiv.org/abs/2502.10050}
\showURL{%
\tempurl}


\bibitem[Pervan(1998)]%
        {PERVAN1998149}
\bibfield{author}{\bibinfo{person}{Graham~P Pervan}.}
  \bibinfo{year}{1998}\natexlab{}.
\newblock \showarticletitle{A review of research in Group Support Systems:
  leaders, approaches and directions}.
\newblock \bibinfo{journal}{\emph{Decision Support Systems}}
  \bibinfo{volume}{23}, \bibinfo{number}{2} (\bibinfo{year}{1998}),
  \bibinfo{pages}{149--159}.
\newblock
\showISSN{0167-9236}
\urldef\tempurl%
\url{https://doi.org/10.1016/S0167-9236(98)00041-4}
\showDOI{\tempurl}


\bibitem[Petruzzelli et~al\mbox{.}(2026)]%
        {Petruzzelli2026Generative}
\bibfield{author}{\bibinfo{person}{Alessandro Petruzzelli},
  \bibinfo{person}{Cataldo Musto}, {and} \bibinfo{person}{Dietmar Jannach}.}
  \bibinfo{year}{2026}\natexlab{}.
\newblock \showarticletitle{Generative User Simulators in Recommendation: A
  Survey}.
\newblock \bibinfo{journal}{\emph{SSRN Electronic Journal}}
  (\bibinfo{year}{2026}).
\newblock
\urldef\tempurl%
\url{https://doi.org/10.2139/ssrn.7269678}
\showDOI{\tempurl}

\bibitem[Ricci and Delić(2025)]%
        {ricci2025wideningrolegrouprecommender}
\bibfield{author}{\bibinfo{person}{Francesco Ricci} {and} \bibinfo{person}{Amra
  Delić}.} \bibinfo{year}{2025}\natexlab{}.
\newblock \bibinfo{title}{Widening the Role of Group Recommender Systems with
  CAJO}.
\newblock
\newblock
\showeprint[arxiv]{2504.05934}~[cs.IR]
\urldef\tempurl%
\url{https://arxiv.org/abs/2504.05934}
\showURL{%
\tempurl}


\bibitem[Rodden(1991)]%
        {Rodden1991Survey}
\bibfield{author}{\bibinfo{person}{Tom Rodden}.}
  \bibinfo{year}{1991}\natexlab{}.
\newblock \showarticletitle{A survey of CSCW systems}.
\newblock \bibinfo{journal}{\emph{Interacting with computers}}
  \bibinfo{volume}{3}, \bibinfo{number}{3} (\bibinfo{year}{1991}),
  \bibinfo{pages}{319--353}.
\newblock


\bibitem[Rook et~al\mbox{.}(2020)]%
        {Rook2020}
\bibfield{author}{\bibinfo{person}{Laurens Rook}, \bibinfo{person}{Adem Sabic},
  {and} \bibinfo{person}{Markus Zanker}.} \bibinfo{year}{2020}\natexlab{}.
\newblock \showarticletitle{Engagement in proactive recommendations}.
\newblock \bibinfo{journal}{\emph{Journal of Intelligent Information Systems}}
  \bibinfo{volume}{54}, \bibinfo{number}{1} (\bibinfo{year}{2020}),
  \bibinfo{pages}{79--100}.
\newblock
\urldef\tempurl%
\url{https://doi.org/10.1007/s10844-018-0529-0}
\showDOI{\tempurl}


\bibitem[Sapkota et~al\mbox{.}(2025)]%
        {Sapkota2025MultiParty}
\bibfield{author}{\bibinfo{person}{Sagar Sapkota},
  \bibinfo{person}{Mohammad~Saqib Hasan}, \bibinfo{person}{Mubarak Shah}, {and}
  \bibinfo{person}{Santu Karmaker}.} \bibinfo{year}{2025}\natexlab{}.
\newblock \showarticletitle{Multi-party conversational agents: A survey}.
\newblock \bibinfo{journal}{\emph{arXiv preprint arXiv:2505.18845}}
  (\bibinfo{year}{2025}).
\newblock


\bibitem[Schedl et~al\mbox{.}(2024)]%
        {schedl2024importance}
\bibfield{author}{\bibinfo{person}{Markus Schedl}, \bibinfo{person}{Oleg
  Lesota}, \bibinfo{person}{Stefan Brandl}, \bibinfo{person}{Mohammad Lotfi},
  \bibinfo{person}{Gustavo Junior~Escobedo Ticona}, {and}
  \bibinfo{person}{Shahed Masoudian}.} \bibinfo{year}{2024}\natexlab{}.
\newblock \bibinfo{title}{The Importance of Cognitive Biases in the
  Recommendation Ecosystem}.
\newblock
\newblock
\showeprint[arxiv]{2408.12492}~[cs.IR]
\urldef\tempurl%
\url{https://arxiv.org/abs/2408.12492}
\showURL{%
\tempurl}


\bibitem[Shi et~al\mbox{.}(2024)]%
        {Shi2024Large}
\bibfield{author}{\bibinfo{person}{Wentao Shi}, \bibinfo{person}{Xiangnan He},
  \bibinfo{person}{Yang Zhang}, \bibinfo{person}{Chongming Gao},
  \bibinfo{person}{Xinyue Li}, \bibinfo{person}{Jizhi Zhang},
  \bibinfo{person}{Qifan Wang}, {and} \bibinfo{person}{Fuli Feng}.}
  \bibinfo{year}{2024}\natexlab{}.
\newblock \showarticletitle{Large Language Models are Learnable Planners for
  Long-Term Recommendation}. In \bibinfo{booktitle}{\emph{Proceedings of the
  47th International ACM SIGIR Conference on Research and Development in
  Information Retrieval}} (Washington DC, USA) \emph{(\bibinfo{series}{SIGIR
  '24})}. \bibinfo{pages}{1893–1903}.
\newblock
\urldef\tempurl%
\url{https://doi.org/10.1145/3626772.3657683}
\showDOI{\tempurl}


\bibitem[Shin et~al\mbox{.}(2023)]%
        {Shin2023IntroBot}
\bibfield{author}{\bibinfo{person}{Donghoon Shin}, \bibinfo{person}{Soomin
  Kim}, \bibinfo{person}{Ruoxi Shang}, \bibinfo{person}{Joonhwan Lee}, {and}
  \bibinfo{person}{Gary Hsieh}.} \bibinfo{year}{2023}\natexlab{}.
\newblock \showarticletitle{IntroBot: Exploring the use of chatbot-assisted
  familiarization in online collaborative groups}. In
  \bibinfo{booktitle}{\emph{Proceedings of the 2023 CHI Conference on Human
  Factors in Computing Systems}}. \bibinfo{pages}{1--13}.
\newblock


\bibitem[Shin et~al\mbox{.}(2021)]%
        {Shin2021BlahBlahBot}
\bibfield{author}{\bibinfo{person}{Donghoon Shin}, \bibinfo{person}{Sangwon
  Yoon}, \bibinfo{person}{Soomin Kim}, {and} \bibinfo{person}{Joonhwan Lee}.}
  \bibinfo{year}{2021}\natexlab{}.
\newblock \showarticletitle{Blahblahbot: Facilitating conversation between
  strangers using a chatbot with ml-infused personalized topic suggestion}. In
  \bibinfo{booktitle}{\emph{Extended Abstracts of the 2021 CHI Conference on
  Human Factors in Computing Systems}}. \bibinfo{pages}{1--6}.
\newblock


\bibitem[Singhal and Pal(2024)]%
        {Singhal2024SOTS}
\bibfield{author}{\bibinfo{person}{Shilpa Singhal} {and}
  \bibinfo{person}{Kunwar Pal}.} \bibinfo{year}{2024}\natexlab{}.
\newblock \showarticletitle{State of art and emerging trends on group
  recommender system: a comprehensive review}.
\newblock \bibinfo{journal}{\emph{International Journal of Multimedia
  Information Retrieval}}  \bibinfo{volume}{13} (\bibinfo{year}{2024}).
\newblock
\urldef\tempurl%
\url{https://doi.org/10.1007/s13735-024-00329-5}
\showDOI{\tempurl}


\bibitem[Stefik et~al\mbox{.}(1987)]%
        {Stefik1987Beyond}
\bibfield{author}{\bibinfo{person}{Mark Stefik}, \bibinfo{person}{Gregg
  Foster}, \bibinfo{person}{Daniel~G Bobrow}, \bibinfo{person}{Kenneth Kahn},
  \bibinfo{person}{Stan Lanning}, {and} \bibinfo{person}{Lucy Suchman}.}
  \bibinfo{year}{1987}\natexlab{}.
\newblock \showarticletitle{Beyond the chalkboard: Computer support for
  collaboration and problem solving in meetings}.
\newblock \bibinfo{journal}{\emph{Commun. ACM}} \bibinfo{volume}{30},
  \bibinfo{number}{1} (\bibinfo{year}{1987}), \bibinfo{pages}{32--47}.
\newblock


\bibitem[Tintarev and Masthoff(2007)]%
        {tintarev2007survey}
\bibfield{author}{\bibinfo{person}{Nava Tintarev} {and} \bibinfo{person}{Judith
  Masthoff}.} \bibinfo{year}{2007}\natexlab{}.
\newblock \showarticletitle{A Survey of Explanations in Recommender Systems}.
  In \bibinfo{booktitle}{\emph{Proceedings of the 3rd International Workshop on
  Web Personalisation, Recommender Systems and Intelligent User Interfaces
  (WPRSIUI'07)}}. \bibinfo{publisher}{IEEE Computer Society},
  \bibinfo{pages}{801--810}.
\newblock
\urldef\tempurl%
\url{https://doi.org/10.1109/ICDEW.2007.4401070}
\showDOI{\tempurl}


\bibitem[Wagne et~al\mbox{.}(2025)]%
        {Wagne2025CRS}
\bibfield{author}{\bibinfo{person}{Ahmadou Wagne}, \bibinfo{person}{Thomas
  Kolb}, \bibinfo{person}{Ashmi Banerjee}, \bibinfo{person}{Fatemeh Nazary},
  \bibinfo{person}{Julia Neidhardt}, {and} \bibinfo{person}{Yashar Deldjoo}.}
  \bibinfo{year}{2025}\natexlab{}.
\newblock \bibinfo{title}{Conversational Recommender Systems Using Generative
  Models (Gen-CRS): A Literature Review}.
\newblock \bibinfo{howpublished}{preprint}.
\newblock
\urldef\tempurl%
\url{https://doi.org/10.13140/RG.2.2.23233.62564}
\showDOI{\tempurl}


\bibitem[Wagner et~al\mbox{.}(2022)]%
        {Wagner2022Moderation}
\bibfield{author}{\bibinfo{person}{Nicolas Wagner}, \bibinfo{person}{Matthias
  Kraus}, \bibinfo{person}{Tibor Tonn}, {and} \bibinfo{person}{Wolfgang
  Minker}.} \bibinfo{year}{2022}\natexlab{}.
\newblock \showarticletitle{Comparing moderation strategies in group chats with
  multi-user chatbots}. In \bibinfo{booktitle}{\emph{Proceedings of the 4th
  Conference on Conversational User Interfaces}}. \bibinfo{pages}{1--4}.
\newblock


\bibitem[Wang et~al\mbox{.}(2024)]%
        {Wang2024ASurvey}
\bibfield{author}{\bibinfo{person}{Lei Wang}, \bibinfo{person}{Chen Ma},
  \bibinfo{person}{Xueyang Feng}, \bibinfo{person}{Zeyu Zhang},
  \bibinfo{person}{Hao Yang}, \bibinfo{person}{Jingsen Zhang},
  \bibinfo{person}{Zhiyuan Chen}, \bibinfo{person}{Jiakai Tang},
  \bibinfo{person}{Xu Chen}, \bibinfo{person}{Yankai Lin},
  \bibinfo{person}{Wayne~Xin Zhao}, \bibinfo{person}{Zhewei Wei}, {and}
  \bibinfo{person}{Jirong Wen}.} \bibinfo{year}{2024}\natexlab{}.
\newblock \showarticletitle{A survey on large language model based autonomous
  agents}.
\newblock \bibinfo{journal}{\emph{Frontiers of Computer Science}}
  \bibinfo{volume}{18}, \bibinfo{number}{6} (\bibinfo{year}{2024}),
  \bibinfo{pages}{186345}.
\newblock
\urldef\tempurl%
\url{https://doi.org/10.1007/s11704-024-40231-1}
\showDOI{\tempurl}


\bibitem[Wu et~al\mbox{.}(2023)]%
        {Wu2023survey}
\bibfield{author}{\bibinfo{person}{Likang Wu}, \bibinfo{person}{Zhi Zheng},
  \bibinfo{person}{Zhaopeng Qiu}, \bibinfo{person}{Hao Wang},
  \bibinfo{person}{Hongchao Gu}, \bibinfo{person}{Tingjia Shen},
  \bibinfo{person}{Chuan Qin}, \bibinfo{person}{Chen Zhu},
  \bibinfo{person}{Hengshu Zhu}, \bibinfo{person}{Qi Liu}, \bibinfo{person}{Hui
  Xiong}, {and} \bibinfo{person}{Enhong Chen}.}
  \bibinfo{year}{2023}\natexlab{}.
\newblock \bibinfo{title}{A Survey on Large Language Models for
  Recommendation}.
\newblock
\newblock
\showeprint[arxiv]{2305.19860}~[cs.IR]


\bibitem[Wu et~al\mbox{.}(2024)]%
        {Wu2024ASurvey}
\bibfield{author}{\bibinfo{person}{Likang Wu}, \bibinfo{person}{Zhi Zheng},
  \bibinfo{person}{Zhaopeng Qiu}, \bibinfo{person}{Hao Wang},
  \bibinfo{person}{Hongchao Gu}, \bibinfo{person}{Tingjia Shen},
  \bibinfo{person}{Chuan Qin}, \bibinfo{person}{Chen Zhu},
  \bibinfo{person}{Hengshu Zhu}, \bibinfo{person}{Qi Liu}, \bibinfo{person}{Hui
  Xiong}, {and} \bibinfo{person}{Enhong Chen}.}
  \bibinfo{year}{2024}\natexlab{}.
\newblock \showarticletitle{A survey on large language models for
  recommendation}.
\newblock \bibinfo{journal}{\emph{World Wide Web}} \bibinfo{volume}{27},
  \bibinfo{number}{5} (\bibinfo{date}{22 Aug} \bibinfo{year}{2024}),
  \bibinfo{pages}{60}.
\newblock
\urldef\tempurl%
\url{https://doi.org/10.1007/s11280-024-01291-2}
\showDOI{\tempurl}


\bibitem[Yan and Gurkan(2023)]%
        {Yan2023itdepends}
\bibfield{author}{\bibinfo{person}{Bei Yan} {and} \bibinfo{person}{Necdet
  Gurkan}.} \bibinfo{year}{2023}\natexlab{}.
\newblock \showarticletitle{It Depends on the Timing: The Ripple Effect of AI
  on Team Decision-Making}. In \bibinfo{booktitle}{\emph{Proceedings of the
  Hawaii International Conference on System Sciences}}.
\newblock


\bibitem[Yehudai et~al\mbox{.}(2025)]%
        {yehudai2025surveyevaluationllmbasedagents}
\bibfield{author}{\bibinfo{person}{Asaf Yehudai}, \bibinfo{person}{Lilach
  Eden}, \bibinfo{person}{Alan Li}, \bibinfo{person}{Guy Uziel},
  \bibinfo{person}{Yilun Zhao}, \bibinfo{person}{Roy Bar-Haim},
  \bibinfo{person}{Arman Cohan}, {and} \bibinfo{person}{Michal
  Shmueli-Scheuer}.} \bibinfo{year}{2025}\natexlab{}.
\newblock \bibinfo{title}{Survey on Evaluation of LLM-based Agents}.
\newblock
\newblock
\showeprint[arxiv]{2503.16416}~[cs.AI]
\urldef\tempurl%
\url{https://arxiv.org/abs/2503.16416}
\showURL{%
\tempurl}


\bibitem[Yeo et~al\mbox{.}(2026)]%
        {Yeo2026GCA}
\bibfield{author}{\bibinfo{person}{ShunYi Yeo}, \bibinfo{person}{Tianyi Zhang},
  \bibinfo{person}{Scott Bateman}, \bibinfo{person}{Gary Hsieh},
  \bibinfo{person}{Young-Ho Kim}, \bibinfo{person}{Simon~Tangi Perrault},
  \bibinfo{person}{Jiannan Li}, {and} \bibinfo{person}{Anthony Tang}.}
  \bibinfo{year}{2026}\natexlab{}.
\newblock \showarticletitle{Group Conversational Agents: A Review of Designs
  that Support and Shape Group Interaction}. In
  \bibinfo{booktitle}{\emph{Proceedings of the 2026 Designing Interactive
  Systems Conference}}. \bibinfo{pages}{2765--2779}.
\newblock


\bibitem[Yu et~al\mbox{.}(2006)]%
        {Yu2006TV}
\bibfield{author}{\bibinfo{person}{Zhiwen Yu}, \bibinfo{person}{Xingshe Zhou},
  \bibinfo{person}{Yanbin Hao}, {and} \bibinfo{person}{Jianhua Gu}.}
  \bibinfo{year}{2006}\natexlab{}.
\newblock \showarticletitle{TV Program Recommendation for Multiple Viewers
  Based on user Profile Merging}.
\newblock \bibinfo{journal}{\emph{User Modeling and User-Adapted Interaction}}
  \bibinfo{volume}{16}, \bibinfo{number}{1} (\bibinfo{year}{2006}),
  \bibinfo{pages}{63–82}.
\newblock
\urldef\tempurl%
\url{https://doi.org/10.1007/s11257-006-9005-6}
\showDOI{\tempurl}


\bibitem[Zhang and Cranshaw(2018)]%
        {Zhang2018MakingSense}
\bibfield{author}{\bibinfo{person}{Amy~X. Zhang} {and} \bibinfo{person}{Justin
  Cranshaw}.} \bibinfo{year}{2018}\natexlab{}.
\newblock \showarticletitle{Making Sense of Group Chat through Collaborative
  Tagging and Summarization}.
\newblock \bibinfo{journal}{\emph{Proc. ACM Hum.-Comput. Interact.}}
  \bibinfo{volume}{2}, \bibinfo{number}{CSCW} (\bibinfo{year}{2018}).
\newblock
\urldef\tempurl%
\url{https://doi.org/10.1145/3274465}
\showDOI{\tempurl}


\bibitem[Zhang et~al\mbox{.}(2025b)]%
        {Zhang2025EvalAgent}
\bibfield{author}{\bibinfo{person}{Guangping Zhang}, \bibinfo{person}{Peng
  Zhang}, \bibinfo{person}{Jiahao Liu}, \bibinfo{person}{Zhuoheng Li},
  \bibinfo{person}{Dongsheng Li}, \bibinfo{person}{Hansu Gu},
  \bibinfo{person}{Tun Lu}, {and} \bibinfo{person}{Ning Gu}.}
  \bibinfo{year}{2025}\natexlab{b}.
\newblock \showarticletitle{EvalAgent: Towards Evaluating News Recommender
  Systems with LLM-based Agents}. In \bibinfo{booktitle}{\emph{Proceedings of
  the 34th ACM International Conference on Information and Knowledge
  Management}} \emph{(\bibinfo{series}{CIKM '25})}.
  \bibinfo{pages}{4086–4095}.
\newblock
\urldef\tempurl%
\url{https://doi.org/10.1145/3746252.3761127}
\showDOI{\tempurl}


\bibitem[Zhang et~al\mbox{.}(2025a)]%
        {cacm25llmagentplatform}
\bibfield{author}{\bibinfo{person}{Jizhi Zhang}, \bibinfo{person}{Keqin Bao},
  \bibinfo{person}{Wenjie Wang}, \bibinfo{person}{Yang Zhang},
  \bibinfo{person}{Wentao Shi}, \bibinfo{person}{Wanhong Xu},
  \bibinfo{person}{Fuli Feng}, {and} \bibinfo{person}{Tat-Seng Chua}.}
  \bibinfo{year}{2025}\natexlab{a}.
\newblock \showarticletitle{Envisioning Recommendations on an LLM-Based Agent
  Platform}.
\newblock \bibinfo{journal}{\emph{Commun. ACM}} \bibinfo{volume}{68},
  \bibinfo{number}{05} (\bibinfo{date}{April} \bibinfo{year}{2025}),
  \bibinfo{pages}{48–57}.
\newblock
\showISSN{0001-0782}
\urldef\tempurl%
\url{https://doi.org/10.1145/3699952}
\showDOI{\tempurl}


\bibitem[Zheng et~al\mbox{.}(2023)]%
        {Zheng2023Competent}
\bibfield{author}{\bibinfo{person}{Chengbo Zheng}, \bibinfo{person}{Yuheng Wu},
  \bibinfo{person}{Chuhan Shi}, \bibinfo{person}{Shuai Ma},
  \bibinfo{person}{Jiehui Luo}, {and} \bibinfo{person}{Xiaojuan Ma}.}
  \bibinfo{year}{2023}\natexlab{}.
\newblock \showarticletitle{Competent but rigid: Identifying the gap in
  empowering AI to participate equally in group decision-making}. In
  \bibinfo{booktitle}{\emph{Proceedings of the 2023 CHI Conference on Human
  Factors in Computing Systems}}. \bibinfo{pages}{1--19}.
\newblock


\bibitem[Zhu et~al\mbox{.}(2025)]%
        {Zhu20225ALLMbased}
\bibfield{author}{\bibinfo{person}{Lixi Zhu}, \bibinfo{person}{Xiaowen Huang},
  {and} \bibinfo{person}{Jitao Sang}.} \bibinfo{year}{2025}\natexlab{}.
\newblock \showarticletitle{A LLM-based Controllable, Scalable, Human-Involved
  User Simulator Framework for Conversational Recommender Systems}. In
  \bibinfo{booktitle}{\emph{Proceedings of the ACM on Web Conference 2025}}
  \emph{(\bibinfo{series}{WWW '25})}. \bibinfo{pages}{4653–4661}.
\newblock
\urldef\tempurl%
\url{https://doi.org/10.1145/3696410.3714858}
\showDOI{\tempurl}


\bibitem[Álvarez Márquez and Ziegler(2018)]%
        {alvarez2018negotiation}
\bibfield{author}{\bibinfo{person}{Jesús Álvarez Márquez} {and}
  \bibinfo{person}{Jürgen Ziegler}.} \bibinfo{year}{2018}\natexlab{}.
\newblock \showarticletitle{Negotiation and Reconciliation of Preferences in a
  Group Recommender System}.
\newblock \bibinfo{journal}{\emph{Journal of Information Processing}}
  \bibinfo{volume}{26} (\bibinfo{date}{02} \bibinfo{year}{2018}),
  \bibinfo{pages}{186--200}.
\newblock
\urldef\tempurl%
\url{https://doi.org/10.2197/ipsjjip.26.186}
\showDOI{\tempurl}


\end{thebibliography}


\end{document}